\documentclass[12pt]{article}
\pdfoutput=1   
  
\usepackage{hyperref}    
\usepackage{amsmath}      
\usepackage{amssymb} 
\usepackage{graphicx}
\usepackage{url}
\usepackage{enumerate}
 
\usepackage{float,wrapfig,color} 

\allowdisplaybreaks
\def\eq#1{(\ref{#1})}
\def\s[#1\s]{\begin{align}\begin{split}#1\end{split}\end{align}}
\def\[#1\]{\begin{align}#1\end{align}}

\begin{document}

\begin{titlepage} 

\title{
\hfill\parbox{4cm}{ \normalsize YITP-22-66}\\  
\vspace{1cm} 
Splitting-merging transitions in tensor-vectors systems \\ in exact large-$N$ limits
}

\author{
Naoki Sasakura\footnote{sasakura@yukawa.kyoto-u.ac.jp}
\\
{\small{\it Yukawa Institute for Theoretical Physics, Kyoto University, }}\\
{\small {\it and } } \\
{\small{\it CGPQI, Yukawa Institute for Theoretical Physics, Kyoto University,}} \\
{\small{\it Kitashirakawa, Sakyo-ku, Kyoto 606-8502, Japan}}
}


\maketitle

\begin{abstract} 
Matrix models have phase transitions in which distributions of variables change topologically like 
the Gross-Witten-Wadia transition.
In a recent study, similar splitting-merging behavior of distributions of dynamical variables was 
observed in a tensor-vectors
system by numerical simulations.
In this paper, we study the system exactly in some large-$N$ limits, in which the distributions are 
discrete sets of configurations rather than continuous.
We find cascades of first-order phase transitions for fixed tensors, and first- and second-order phase transitions for random
tensors, being characterized by breaking patterns of replica symmetries. 
The system is of interest across three different subjects at least:
The splitting dynamics plays essential roles in emergence of classical spacetimes in a tensor model
of quantum gravity; 
The splitting dynamics automatically detects the rank of a tensor in the tensor rank decomposition in data analysis; 
The system provides a variant of the spherical $p$-spin model for spin glasses with a new 
non-trivial parameter. 
We discuss some implications of the results from these perspectives.
The results are compared with some numerical simulations to check the large-$N$ convergence and the assumptions made
in the analysis.
\end{abstract}
\end{titlepage}

\section{Introduction}
\label{sec:introduction}
The purpose of this paper is to better understand the dynamics of the dynamical system defined by the partition function,
\[ 
Z_{N,R}(\beta, C)= \int_{I} d\phi\, e^{-\beta(C-\phi\phi\phi)^2}, 
\label{eq:system}
\]
where $\beta$ is an inverse temperature, $C$ denotes a real symmetric tensor of order three, 
$C_{abc}=C_{bac}=C_{bca}\ (a,b,c=1,2,\ldots,N)$,  
the integration variables are $\phi_a^i\ (a=1,2,\ldots,N,\ i=1,2,\ldots,R)$, the integration region is 
$I=\mathbb{R}^{NR}$, the integration measure is $d\phi=\prod_{a=1}^N \prod_{i=1}^R d\phi_a^i$,
and we use a short-hand notation,
\[
(C-\phi\phi\phi)^2=\left(C_{abc}-\sum_{i=1}^R \phi_a^i\phi_b^i\phi_c^i\right)\left(
C_{abc}-\sum_{i=1}^R \phi_a^i\phi_b^i\phi_c^i\right),
\]
where pairwise repeated lower indices are assumed to be summed over, as is assumed throughout this paper.
On the other hand summations over the upper indices have to be always explicitly indicated.
Note that, in addition to an $O(N)$ symmetry with respect to the lower indices,
 the system \eq{eq:system} is invariant under relabeling the upper index of $\phi_a^i$ 
(namely, invariant under $\phi^i_a \leftrightarrow \phi^j_a$).
In this paper, we call it {\it real} replica symmetry, distinguishing it from the replica symmetry which appears later
in the replica trick.

One of the motivations to study the system \eq{eq:system} comes from a tensor model in the Hamiltonian formalism,
which we call canonical tensor model (CTM) \cite{Sasakura:2011sq,Sasakura:2012fb}.
Tensor models were originally introduced as a generalization of the matrix models, which are successful in 
describing two-dimensional quantum gravity, with a hope to extend the success to higher
 dimensions \cite{Ambjorn:1990ge,Sasakura:1990fs,Godfrey:1990dt,Gurau:2009tw}. However,
these tensor models do not generate macroscopic spacetimes, 
suffering from dominance of singular objects like branched polymers \cite{Bonzom:2011zz,Gurau:2011xp}.
A motivation of considering CTM is to overcome the issue
by introducing a temporal direction into tensor models, trying to follow the
success of the causal dynamical triangulation over the dynamical triangulation
in generating macroscopic spacetimes \cite{Ambjorn:2004qm}, 
where the former has a temporal direction, while the latter not. 
Indeed, in \cite{Kawano:2021vqc}, it was explicitly shown by numerical simulations that a wave function of CTM 
seems to have a twofold phase structure, and classical spacetimes emerge in the one which we call the classical 
phase.\footnote{The other was called 
the quantum phase.}
Here the transition between the two phases can be characterized by splitting-merging transitions
of distributions of the dynamical variables, like those in the matrix counterparts, such as 
the Gross-Witten-Wadia transition \cite{Gross:1980he,Wadia:1980cp} 
and the transitions among multi-cut large-$N$ solutions \cite{Eynard:2016yaa}.
However, the simulations were not convincing enough 
to conclude whether the phases are really different (in some large-$N$ limits),  
not just separated by crossovers.
The system \eq{eq:system} is a part of this wave function\footnote{The wave function has an integral expression, which 
is a multi-variable extension of the integral representation of Airy functions. Its integrand is complex, but, by taking only the modus of the integrand, one obtains the system \eq{eq:system}. 
In \cite{Kawano:2021vqc} the wave function, of complex values, was analyzed by the re-weighting method
of the Monte Carlo simulations, in which the system \eq{eq:system} played the role 
of a statistical system with a positive weight in the method.},
and plays essential roles in the dynamics of the above transition.
A result of this paper is that, at least in the large-$N$ limits we consider, 
the system \eq{eq:system} has  sharp phase transitions characterized by breaking patterns of the 
real/genuine replica symmetries.
The phases which appear at large values of $\beta$ correspond to the classical 
phase found in the previous paper.\footnote{
In fact, we will find cascades of first-order phase transitions in this paper, meaning that what was called the classical 
phase in \cite{Kawano:2021vqc} could be a collection of phases in general. This, however, must be taken with caution,
since the large-$N$ limits we take in this paper is different from that in the tensor model, which should be $R\propto N^2$
(See the last section).}

Another motivation of studying \eq{eq:system} comes from that the exponent in \eq{eq:system} 
can be used as a cost function of the tensor rank
decomposition \cite{SAPM:SAPM192761164,Carroll1970,Landsberg2012,comon:hal-00923279}, 
which is an important technique in data analysis.\footnote{See \cite{Ouerfelli:2022rus} for some
developments to data analysis techniques from random tensor studies.} 
In the present case of a real symmetric tensor $C$ of order three, a real symmetric tensor rank
decomposition is defined to find $\phi_a^i \in \mathbb{R}$ satsfying
\[
C_{abc}=\sum_{i=1}^R \phi_a^i \phi_b^i \phi_c^i,
\label{eq:decomp}
\]
which is equivalent to require the cost function to vanish.
The minimum value of $R$ which realizes this decomposition for a $C$
is called the rank of $C$.
The decomposition \eq{eq:decomp} is a sort of an extension of the singular value decomposition 
of the matrix to the tensor, 
but the hardness is largely different \cite{nphard}. 
While a matrix can be decomposed by straightforward procedures, 
a practical method for the tensor case is to optimize $\phi_a^i$ so that the cost function $(C-\phi\phi\phi)^2$ 
be minimized or vanish \cite{hack}.
Here one of the difficult issues is that we do not have prior knowledge of an appropriate value of  $R$ for a given $C$: 
If we take a too large $R$ for the optimization, the decomposition will be overfitting, 
and, if a too small $R$ is taken, we will miss some properties of $C$. 
What was found in \cite{Kawano:2021vqc} and is interesting 
in the system \eq{eq:system} is that, when $\beta$ is taken large enough, the dominant configurations 
of $\phi_a^i$ are such that they are separated into two parts, the dominant and minor 
parts\footnote{Here the upper index of $\phi_a^i$ has been relabeled without loss of generality.},
\s[
C_{abc}&=\sum_{i=1}^{R_c} \phi_a^i \phi_b^i \phi_c^i+\sum_{i=R_c+1}^R \phi_a^i \phi_b^i \phi_c^i, \\
\phi_a^i&\not\sim 0\ (i=1,2,\ldots,R_c), \ \ \  
\phi_a^i\sim 0\ (i=R_c+1,R_c+2,\ldots,R).
\label{eq:decoupling}
\s] 
In fact, in the examples considered in \cite{Kawano:2021vqc}, 
the values of $R_c$ agreed with the ranks of $C$ (or very near values in a few large system cases).  
Therefore the system \eq{eq:system} seems to have an intrinsic dynamics which automatically 
detects an appropriate rank for a given $C$. A purpose of this paper is to study this interesting
property, which was found in the previous numerical simulation, by an exact method.
In the long run, understanding the system \eq{eq:system} would provide some solutions 
to the long-standing issues in the tensor rank decomposition.

Another interesting link of the system \eq{eq:system} to physics is spin glasses. It can be regarded 
as a variant of spherical $p$-spin model for spin glasses \cite{pspin,pedestrians}, 
which is defined by the following Hamiltonian
and a constraint,
\s[
&H=C_{a_1a_2\cdots a_p} \phi_{a_1}\phi_{a_2}\cdots\phi_{a_p}, \\
&\phi_a\phi_a=N,
\label{eq:pspin}
\s]
where the tensor $C_{a_1a_2\cdots a_p}$ is assumed to take random 
numbers\footnote{The simplest realization would be a normal distribution.}. 
Thus the system \eq{eq:system} can be regarded as a multi-real-replica extension of the spherical $p$-spin 
model. Note that we are interested in finite $R$, while a replica number is taken to vanish in the replica trick.
Considering the connection to the tensor rank decomposition, it should be a non-trivial question how \eq{eq:system} 
behaves in $R$.

The non-triviality of the $R$-dependence can also be seen in another way.
Note that our system \eq{eq:system} does not have the constraint in \eq{eq:pspin}, 
which prohibits $\phi_a^i$ to run away to infinity, assuring the stability of the model. 
Though the exponent in \eq{eq:system} is semidefinite in our case, 
 it is a non-trivial question whether \eq{eq:system} is finite or not, because 
 the exponent contains flat directions, such as $\phi_a^i=-\phi_a^j$, which extends to infinity.  
This question about the finiteness was systematically studied mainly by numerical methods in 
\cite{Obster:2021xtb}\footnote{There are also some closely related studies \cite{Lionni:2019rty,Sasakura:2019hql,Obster:2020vfo}.}, 
and it was checked/conjectured that the system is finite only for $R\lesssim (N+1)(N+2)/2$.
In this paper, we are free from this instability, because we consider large-$N$ limits with finite $R$. 

This paper is organized as follows.
In Section~\ref{sec:givenc}, we introduce a large-$N$ limit with finite $R$ of the system \eq{eq:system}
for fixed $C$, and derive an exact expression of the free energy in the limit.
In Section~\ref{sec:excase1}, we consider a few representative examples of $C$, and 
study the free energy by explicitly
computing minima of the expression obtained in Section~\ref{sec:givenc}.
As $\beta$ is increased, the system undergoes a cascade of first-order phase transitions, 
in which the number of non-zero $\psi^i$, which appear later, increases one by one, eventually reaches a phase
in which the number agrees with the tensor rank of $C$, and stays there.
In Section~\ref{sec:randomC}, we consider the case of random $C$. 
To incorporate the random $C$, we employ the replica trick up to one-step replica symmetry breaking (1RSB) 
as in the case of the spherical $p$-spin model. 
When $\beta$ is small, the system is in the replica symmetric phase.
As $\beta$ is increased, the system undergoes a first- or second-order phase transition, depending on
a parameter we introduce, and enters the 1RSB spin-glass phase. 
In Section~\ref{sec:comparison}, we perform some numerical simulations and compare with the exact results.

\section{Large-$N$ limits with finite $R$ and fixed $C$}
\label{sec:givenc}
The strategy of this section is that we dip $C$ of a finite dimension into a large-$N$ system. 
More precisely, let us introduce a parameter $n$, 
and assume that $C$ takes non-zero values only within this sub-dimension:
\s[
&C_{a_1a_2a_3}\neq 0,\ \hbox{only if}\ \forall a_i \in \{1,2,\ldots,n\}, \\
& C_{a_1a_2a_3}=0, \hbox{otherwise}.
\label{eq:rangec}
\s]
We assume that $n$ is kept finite in the large-$N$ limit 
(However, we will later consider $n\propto N$ in the large-$N$ limit for random $C$).
Let us introduce new variables to separate $\phi_a^i$ into two parts:
\s[
&\psi_a^i=\phi_a^i \ (a=1,2,\ldots,n), \\
&\tilde \psi_a^i=\phi_a^i\ (a=n+1,n+2,\ldots,N),
\label{eq:division}
\s]
for $\forall i=1,2,\ldots,R$.
With these variables, the exponent $\beta (C-\phi\phi\phi)^2$ of \eq{eq:system} can be rewritten as
\s[
S_{\psi\tilde \psi}(\beta,C)
&=\beta (C-\psi\psi\psi)^2 +3\beta \sum_{i,j=1}^R (\psi^i\cdot \psi^j)^2 \,\tilde \psi^i\cdot\tilde \psi^j \\
&\ \ \ +3\beta\sum_{i,j=1}^R \psi^i\cdot \psi^j \,(\tilde \psi^i\cdot\tilde \psi^j)^2
+\beta\sum_{i,j=1}^R (\tilde \psi^i\cdot\tilde \psi^j)^3,
\s] 
where $\cdot$ denotes the inner product, $\psi^i\cdot \psi^j=\psi_a^i \psi_a^j$, and $C$ abusively 
denotes the $n$-sub-dimensional part.  
$S_{\psi\tilde \psi}(\beta,C)$ is invariant under the $SO(N-n)$ transformation
with respect to the lower index of $\tilde\psi_a^i$, and we can factor out the degrees of freedom by 
introducing the following new variable, which is called overlap in spin glass theory \cite{pedestrians},
\[
\tilde Q_{ij}=\tilde \psi^i \cdot \tilde \psi^j,\ (i,j=1,2,\ldots,R).
\label{eq:tildeQ}
\]
This constraint can be embedded into the system by rewriting the partition function \eq{eq:system} as
\[
Z_{N,R}(\beta,C)=\int d\psi d\tilde \psi d\lambda d\tilde Q\, e^{-S_{\psi\tilde Q}(\beta,C)+ i \sum_{i,j=1}^R \lambda_{ij} 
(\tilde Q_{ij}-\tilde \psi^i \cdot \tilde \psi^j)},
\label{eq:zpsipsitilde}
\]
where we have ignored an irrelevant overall factor, and 
\s[
S_{\psi \tilde Q}(\beta,C)=\beta (C-\psi\psi\psi)^2 +3\beta \sum_{i,j=1}^R (\psi^i\cdot \psi^j)^2 \tilde Q_{ij} 
+3\beta\sum_{i,j=1}^R \psi^i\cdot \psi^j \,(\tilde Q_{ij})^2
+\beta\sum_{i,j=1}^R (\tilde Q_{ij})^3.
\s]
Integrating over $\tilde \psi$ in\eq{eq:zpsipsitilde}
generates a new term $-(N-n)/2\ \log \mathrm{det} \lambda$ in the exponent, and then 
by assuming large-$N$ and carrying out the $\lambda$ integration by taking the saddle point, we obtain
\[
S^{\rm eff}_{\psi\tilde Q}(\bar \beta,C)=(N-n) \left(S_{\psi\tilde Q}(\bar \beta,C)-\frac{1}{2} \ln \mathrm{det} \tilde Q
\right).
\label{eq:seff}
\]
where we have introduced $\bar \beta$ by $\beta=(N-n)\,\bar \beta$. 
Therefore, in the large-$N$ limit, the free energy of the system is given by
\s[
\bar \beta F(\bar \beta,C)&=-\lim_{N\rightarrow \infty} \frac{1}{N} \log Z(\beta,C) \\
&=\min_{\psi, \tilde Q \geq 0} \left(S_{\psi\tilde Q}(\bar \beta,C)-\frac{1}{2} \ln \mathrm{det} \tilde Q
\right),
\label{eq:free}
\s]
where $\tilde Q \geq 0$ represents that $\tilde Q$ is constrained to be a positive semidefinite 
matrix\footnote{Namely, all the eigenvalues are zero or positive.} due to \eq{eq:tildeQ}. 

In general, when the parameters $\bar \beta$ and $C$ are gradually changed, 
the free energy \eq{eq:free} will undergo various first-order phase transitions with finite jumps of the minimum
in $\psi,\tilde Q$. 
As will explicitly be shown in some representative cases in Section~\ref{sec:excase1},
the splitting between the dominant and minor parts \eq{eq:decoupling} occurs
as the results of the first-order phase transitions.

\section{Some representative examples for fixed $C$}
\label{sec:excase1}
In this section, we explicitly compute \eq{eq:free} for some $C$'s as a function of $\bar \beta$.
The minimization must be performed under the condition that $\tilde Q$ is a positive semidefinite
matrix.
A convenient way to implement it is to parameterize $\tilde Q$ as
\[
\tilde Q_{ij}=v^{i}\cdot v^j,
\]
where $v^{i}\ (i=1,2,\ldots,R)$ are $R$-dimensional real vectors which have
non-zero elements only in a triangular part:
\[
v^i_a
\left\{
\begin{array}{l}
\neq 0, \ \hbox{only if } 1\leq a\leq i, \\
=0, \ \hbox{for } i<a\leq R.
\end{array}
\right.
\]
Unless $n$ and $R$ are very large,  the minimum \eq{eq:free} can be obtained 
by repeating the minimization process many times enough, starting from random values of 
$\psi$ and $v$.

\subsection{$n=1$}
Though $C=C_{111}$ cannot even be called a tensor for $n=1$, this simplest case still shows 
the splitting \eq{eq:decoupling}, as we will see below.
For $n=1$, the ``tensor rank decomposition" \eq{eq:decomp} is just a scalar equation,
\[
C=\sum_{i=1}^R (\psi^i)^3,
\]
and,  for $R>1$,  it has a continuously infinite number of solutions extending to infinity.
Considering the abundance of the solutions, 
the dominance shown below of the particular splitted configurations \eq{eq:decomp} in the system \eq{eq:system}
is rather surprising. This dominance should be a non-trivial effect of the integration volume.

\begin{figure}
\begin{center}
\includegraphics[width=4cm]{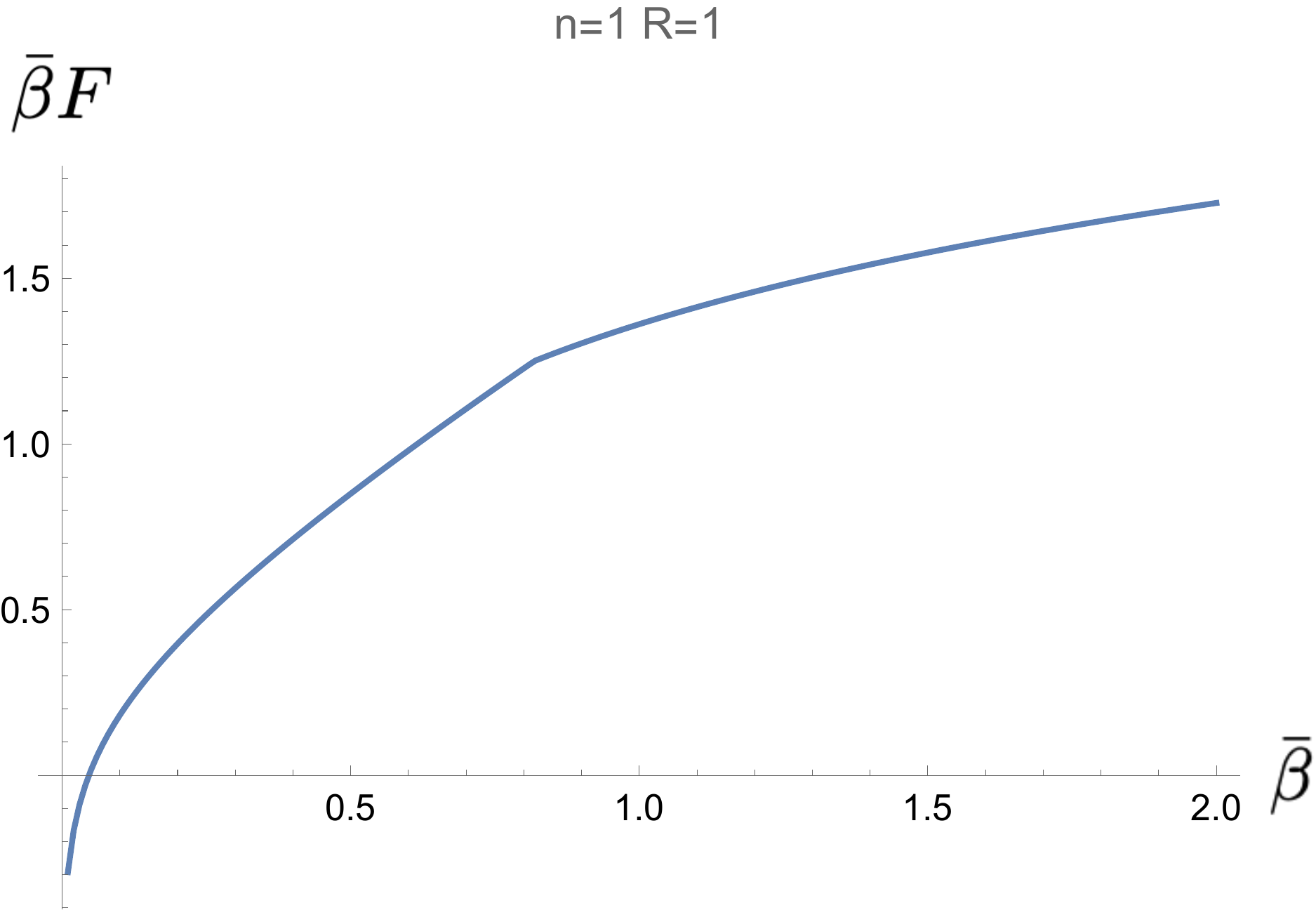}
\hfil
\includegraphics[width=4cm]{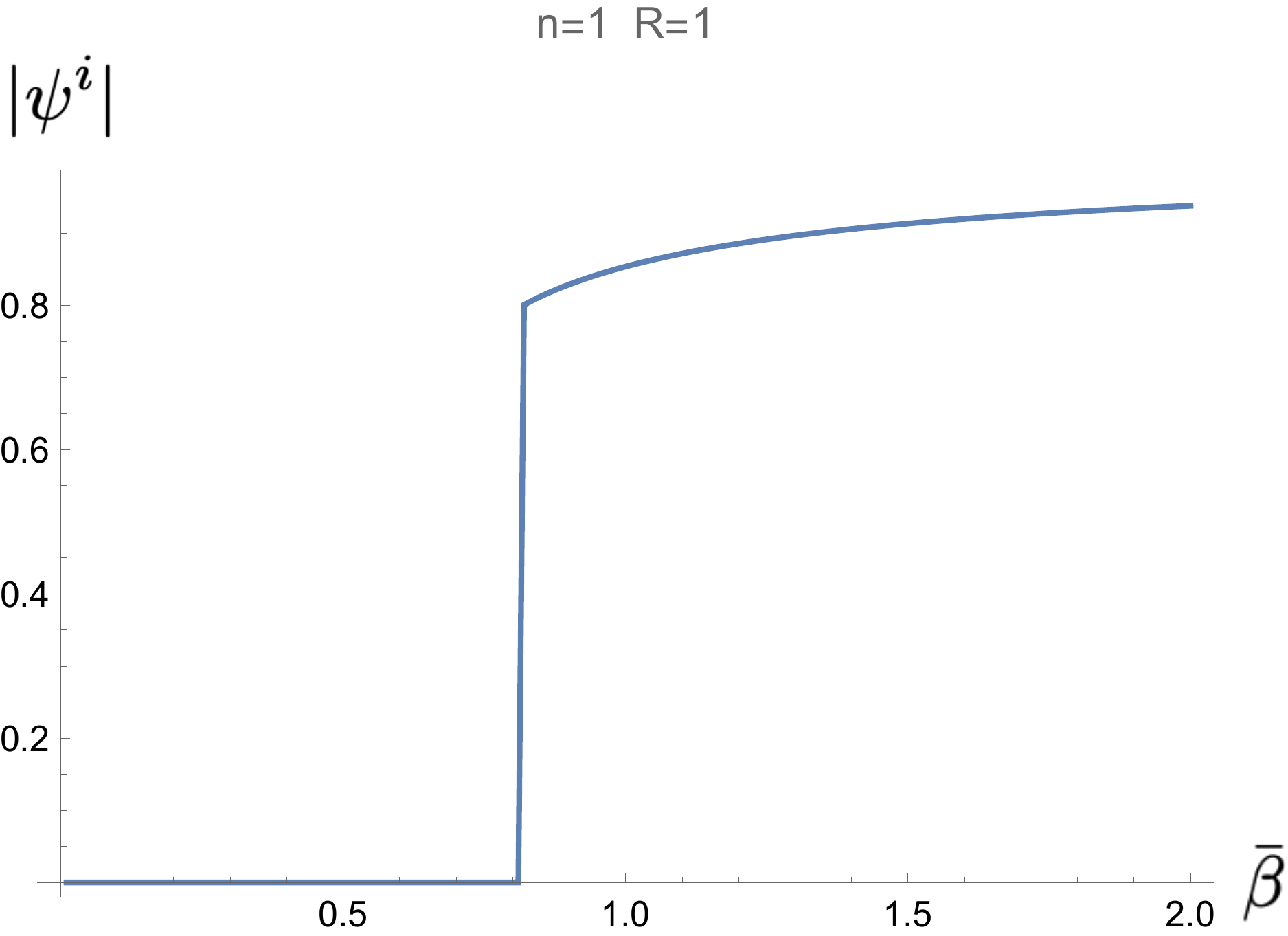}
\hfil
\includegraphics[width=4cm]{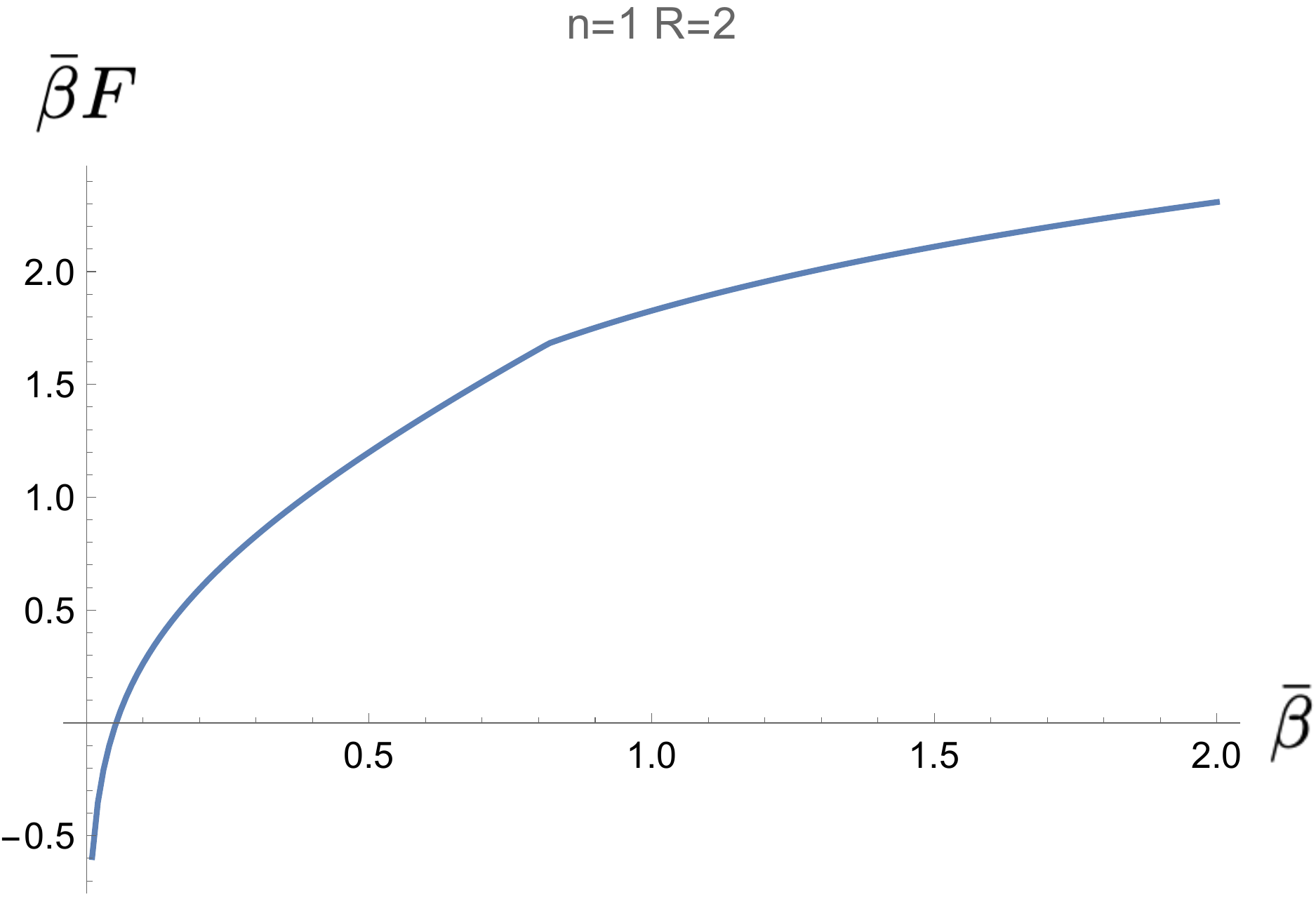}
\hfil
\includegraphics[width=4cm]{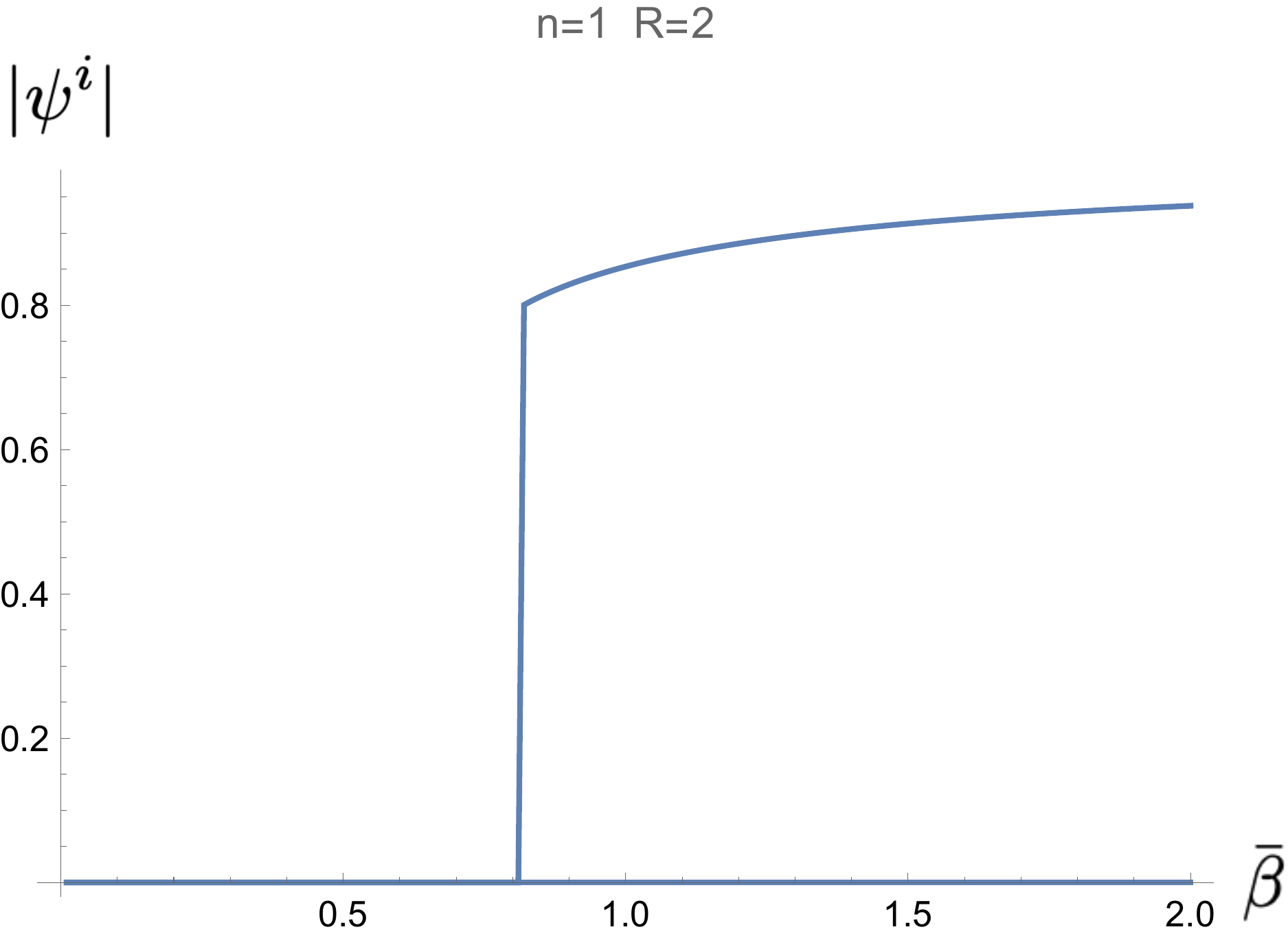}
\caption{Left two panels: $n=1,\ R=1,\ C=1$ are taken. The left of them plots the free energy \eq{eq:free}
against $\bar \beta$. 
There is a first-order phase transition at $\bar \beta\sim 0.8$. The right of them plots the value of $\psi$, which 
has a jump from zero to a finite value.
The right two panels: Similar plots for $n=1,\ R=2,\ C=1$.
Only one of $\psi^1, \psi^2$ gets finite at $\bar \beta > \bar \beta_c$.}
\label{fig:N1}
\end{center}
\end{figure}

Two examples with $R=1,2$ each are shown in Figure~\ref{fig:N1}.  In both cases, there are first-order 
phase transitions at $\bar \beta=\bar \beta_c \sim0.8$. 
In the left example with $R=1$, the only $\psi^1$ makes a jump from zero to a finite value, while, 
in the right example with $R=2$, only one of the $\psi^i$, say $\psi^1$, makes a jump, but the other, $\psi^2$, stays zero:
The splitting \eq{eq:decomp} is realized in the phase at $\bar \beta > \bar \beta_c$. Note that
$\psi^2$, corresponding to the minor part in \eq{eq:decoupling}, is exactly zero 
in this large-$N$ limit. 
One can also check that the above phenomenon is universal for any $R\geq2$: Only one of $\psi^i$ gets non-zero values
and the others stay zero  at $\bar \beta > \bar \beta_c$. 

The above transition can be restated as a real replica symmetry breaking. In the phase at $\bar \beta <\bar \beta_c$, 
the real replica symmetry, namely, the symmetric group $S_R$ interchanging $\phi^i\ (i=1,2,\ldots,R)$,
is unbroken because of $\forall \psi^i=0$. In the other phase at  $\bar \beta >\bar \beta_c$, it is spontaneously broken,
\[
S_R \rightarrow S_{R-1}, \ \hbox{at } \bar \beta=\bar \beta_c,
\]
because one of $\psi^i$ gets finite.

The same symmetric structure can also be checked for $\tilde Q$.

\begin{figure}
\begin{center}
\includegraphics[width=4cm]{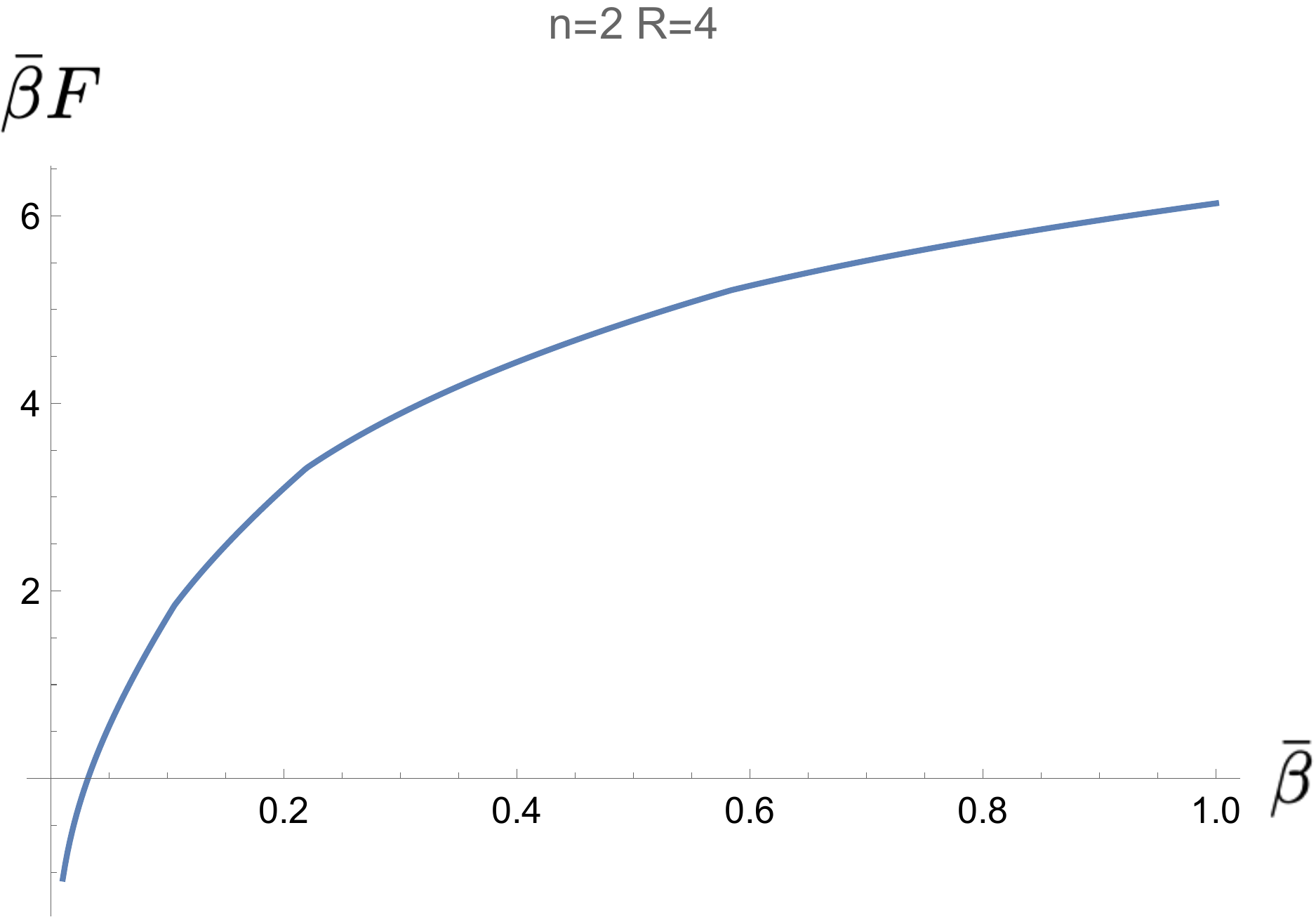}
\hfil
\includegraphics[width=4cm]{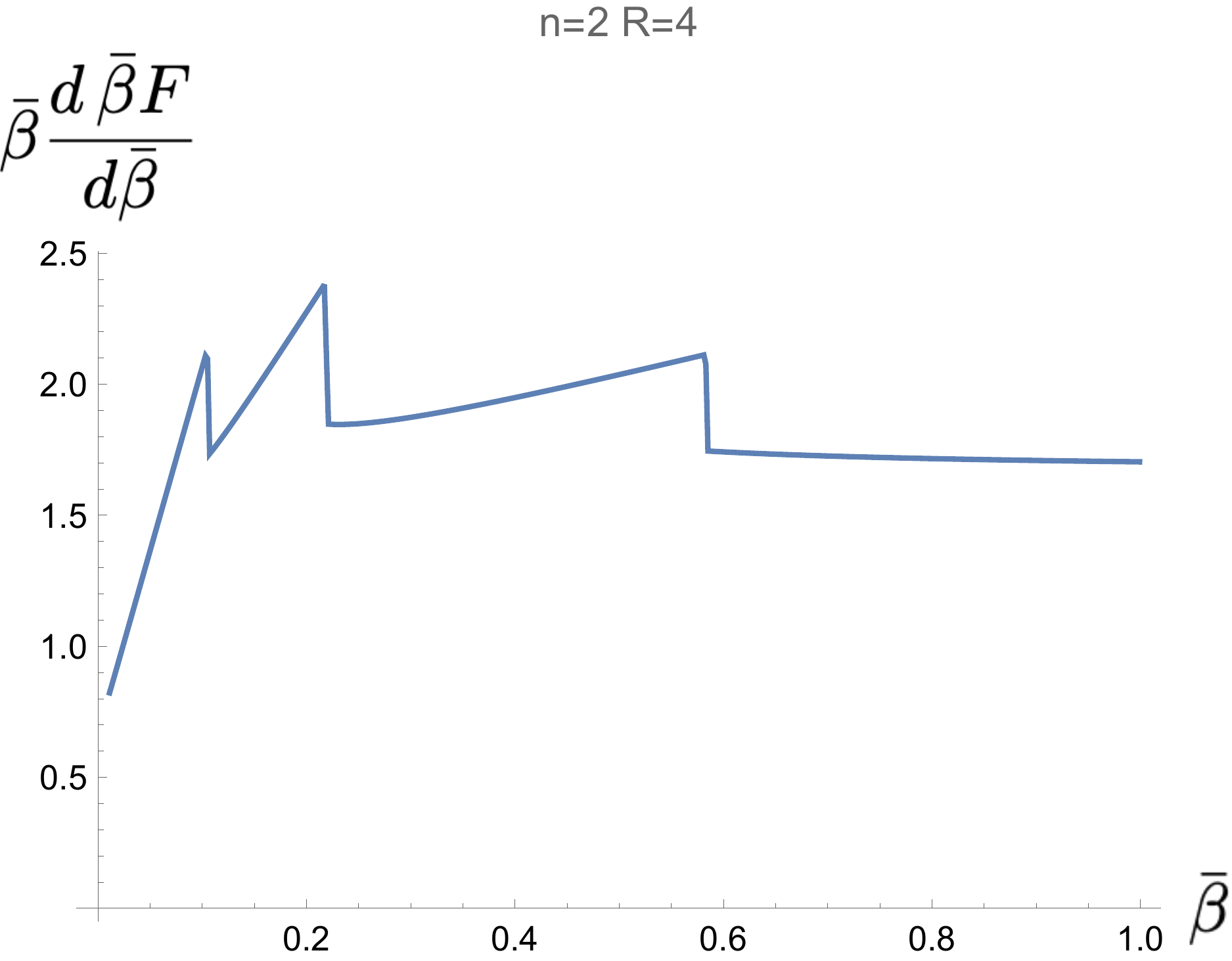}
\hfil
\includegraphics[width=4cm]{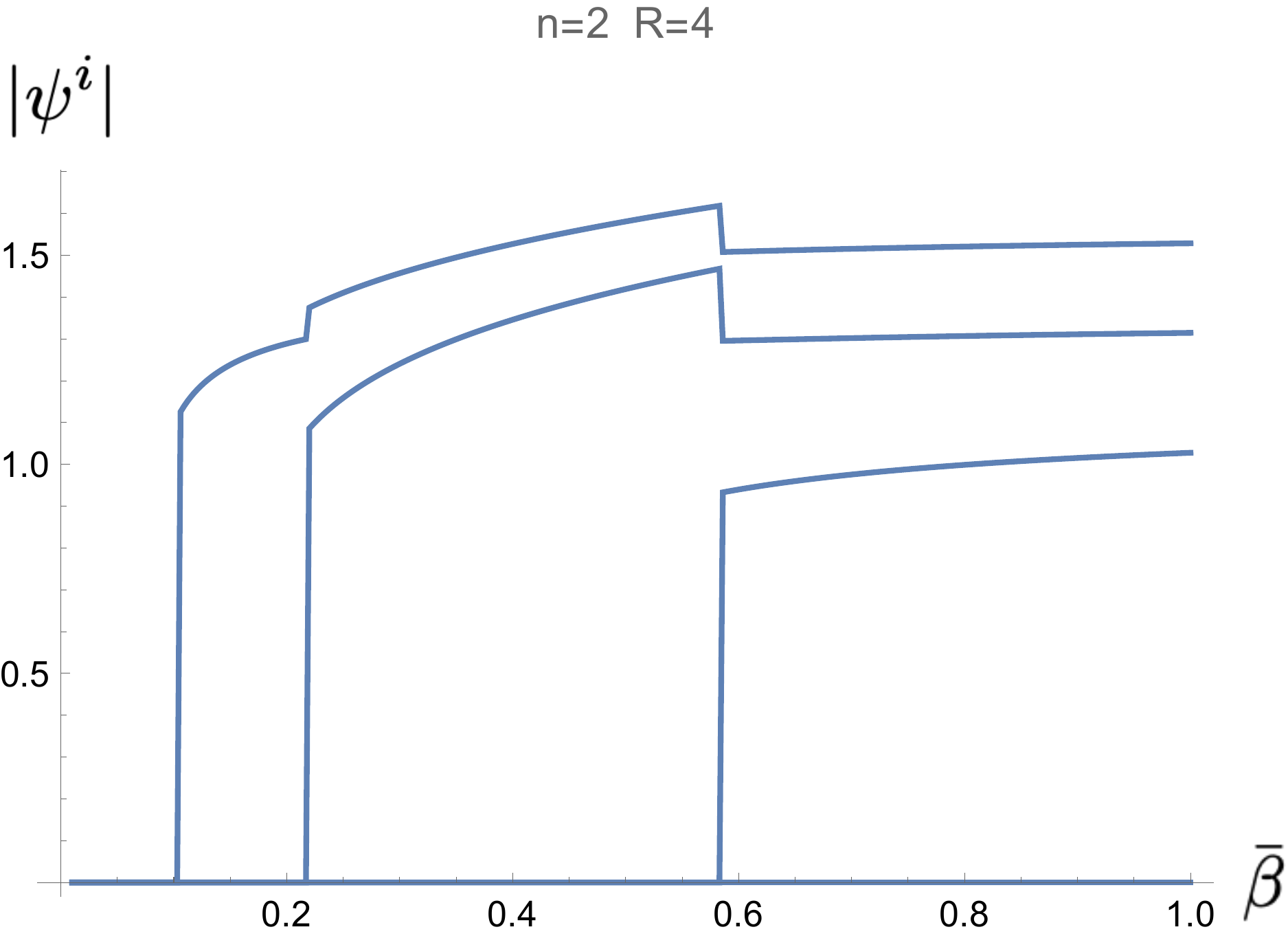}
\caption{An example of $n=2, \ R=4$. $C_{111}=C_{222}=1,\ C_{122}=-2$ is taken for $C$. From the left to the right panels, the free energy \eq{eq:free}, $\bar \beta d\bar \beta F/d\bar \beta$, and $|\psi^i|=\sqrt{\psi^i_a \psi_a^i}$ are 
plotted against $\bar \beta$, respectively. 
}
\label{fig:N2}
\end{center}
\end{figure}

\subsection{$n=2$}
\label{sec:neq2fixedc}
The typical ranks\footnote{The space of $C$ (with a normalization) can be classified by the 
ranks of $C$. Typical ranks are the ranks which appear with a finite measure in the space.
In other words,  when a $C$ is randomly chosen,  the rank of $C$ will be one of the typical ranks, each appearing 
with a non-zero probability.}
 of the $n=2$ real symmetric tensors are 2 and 3 \cite{comon}. 
Therefore, depending on 
the choice of $C$, we typically
find 2 or 3 first-order phase transition points for $R\geq 3$, as $\beta$ is changed.
Figure~\ref{fig:N2} shows an example with $\hbox{rank}(C)=3$ and $R=4$. We indeed 
find three first-order phase transitions. 
The number of non-zero vectors of $\psi^i$ increases one by one from zero to three, as the system undergoes 
each first-order phase transition when $\bar \beta$ is increased. 
One vector keeps vanishing how large $\bar \beta$ is taken, because $R-\hbox{rank}(C)=1$ in this case.
One can check that the splitting, as in \eq{eq:decomp}, is universal for any $R\geq 4$ when $\bar \beta$ 
is taken large enough. In this phase,  the number $\hbox{rank}(C)$ of $\psi^i$ 
takes non-zero values, while the others are exactly zero.  
Thus there are typically the following two possible 
patterns of real replica symmetry breaking for $n=2$:
\s[
&S_R\rightarrow S_{R-1}\rightarrow S_{R-2}, \hbox{ for rank}(C)=2,   \\
&S_R\rightarrow S_{R-1}\rightarrow S_{R-2}\rightarrow S_{R-3}, \hbox{ for rank}(C)=3.
\s] 

\begin{figure}
\begin{center}
\includegraphics[width=7cm]{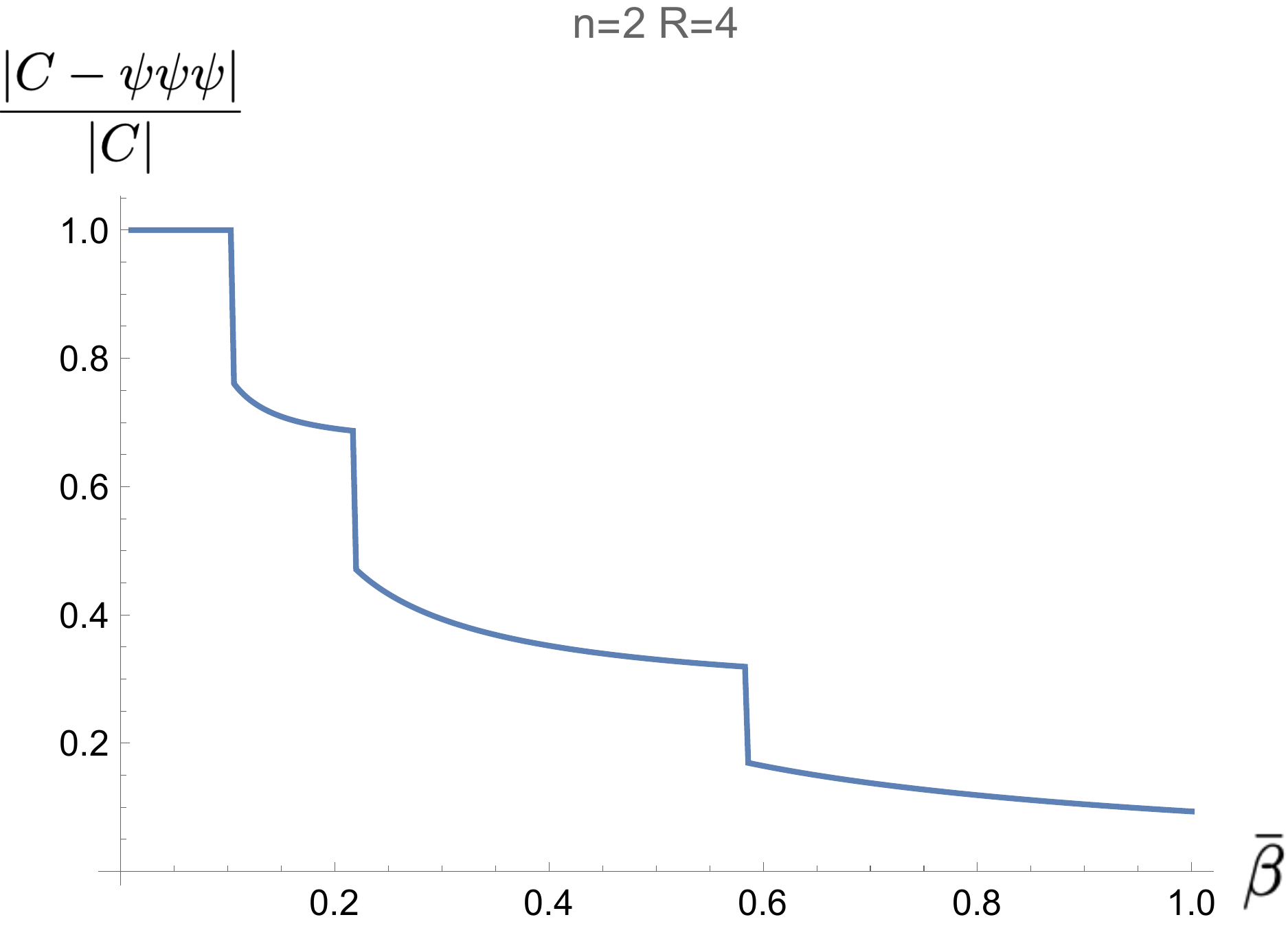}
\caption{ The error of the tensor rank decomposition, $|C-\psi \psi \psi|/|C|$,
is plotted against $\bar \beta$ for the same example as in Figure~\ref{fig:N2}.  }
\label{fig:n2error}
\end{center}
\end{figure}

It would be instructive to see how the tensor rank decomposition of $C$ is performed as $\bar \beta$ is increased.
As shown in Figure~\ref{fig:n2error}, the tensor rank decomposition is improved at each time the 
first-order phase transitions occurs.

\subsection{Lie-group symmetric $C$}
\label{sec:lie}

\begin{figure}
\begin{center}
\includegraphics[width=4cm]{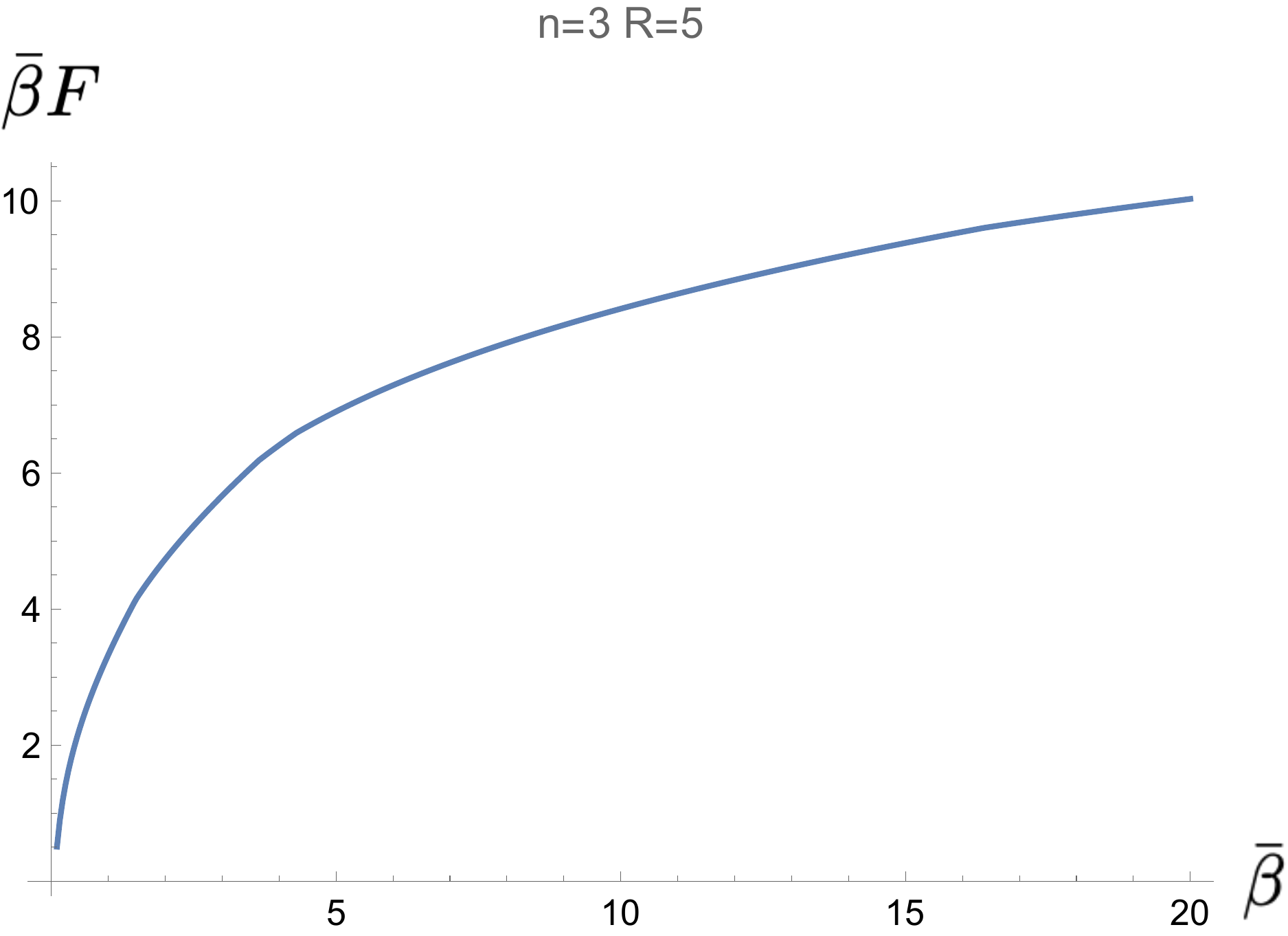}
\hfil
\includegraphics[width=4cm]{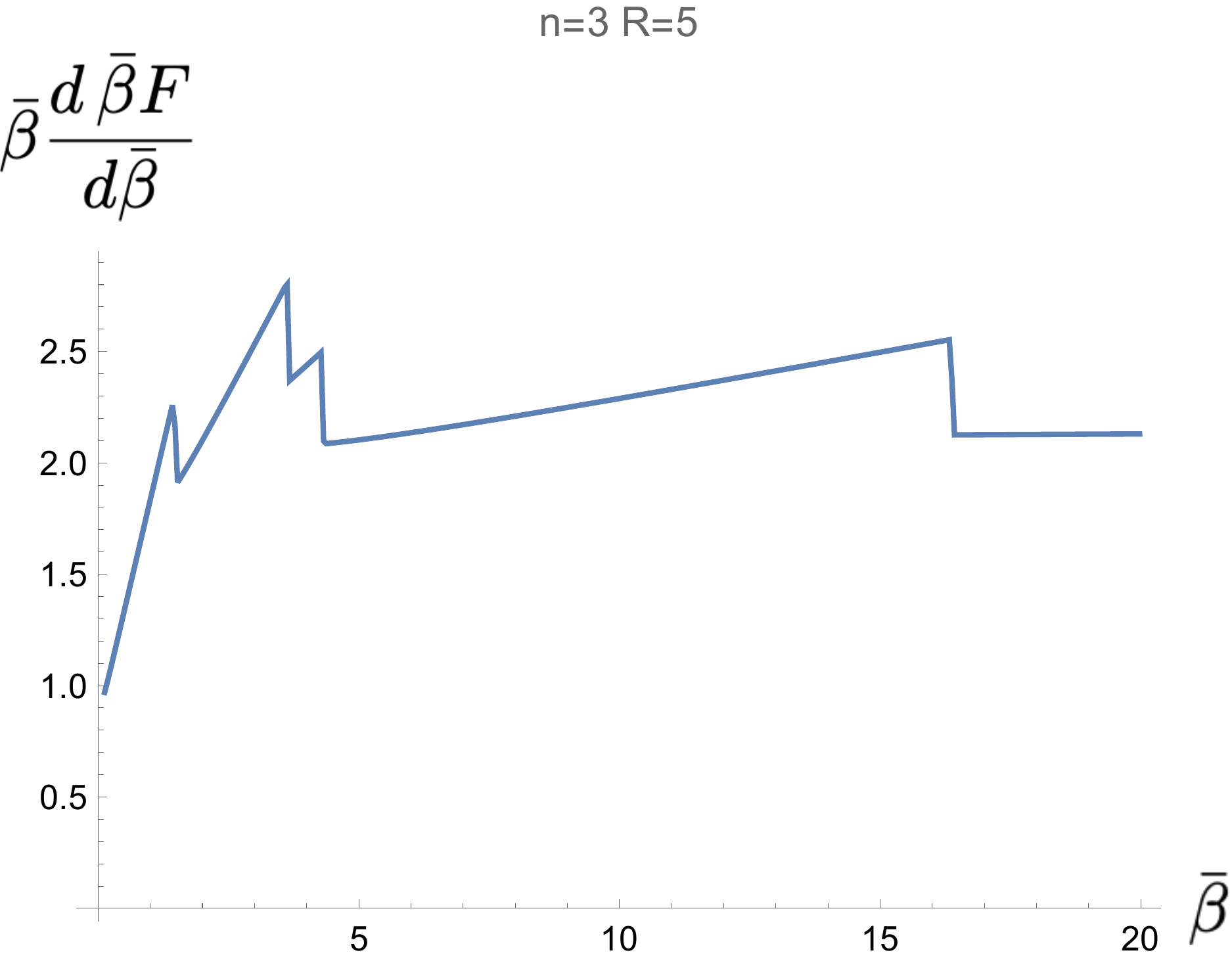}
\hfil
\includegraphics[width=4cm]{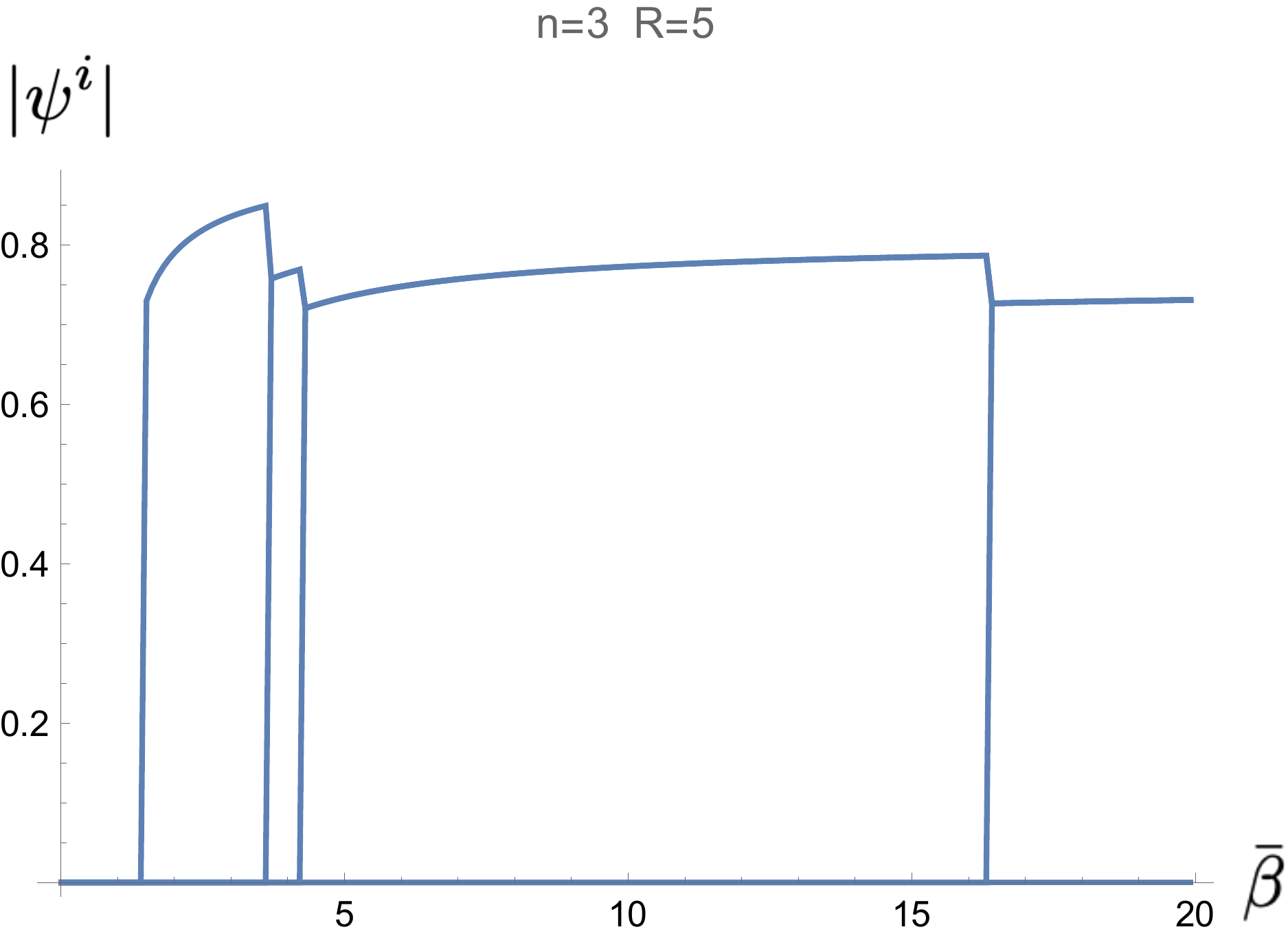}
\caption{An example with $n=3, \ R=5$ and an $SO(2)$ invariant $C$ with $\alpha=0.5$ in \eq{eq:invc}. 
In the right panel, all the non-zero $\psi^i$ have the same size, and are degenerate in the plot.
}
\label{fig:N3}
\end{center}
\end{figure}

For $n\geq 3$, $C$ can be taken Lie-group symmetric. 
As an illustrative example, let us consider the simplest case of $n=3$ and $SO(2)$-invariant $C$.
More precisely, $C$ is given by
\[
C_{abc}=\hbox{const.}\, e^{-\alpha (m_a^2+m_b^2+m_c^2)} \int_0^{2 \pi} d\theta \, f_a f_b f_c,
\label{eq:invc}
\]
where 
\[
f_1=\frac{1}{\sqrt{2}},\ f_2=\cos(\theta),\ f_3=\sin(\theta),
\]
$m_a$ are the angular frequencies of the functions $f_a$, namely, $m_1=0,m_{2}=m_{3}=1$.
The parameter $\alpha$ has been introduced to make smooth the sharp frequency cutoff, at $m=1$ in this case,
by choosing $\alpha\sim O(1)$.
The overall factor $\hbox{const.}$ is a normalization factor for $C_{abc}C_{abc}=1$.
It is easy to check that $C$ is invariant under an arbitrary $SO(2)$ rotation between $f_2$ and $f_3$, which 
corresponds to a shift of $\theta$. 

Figure~\ref{fig:N3} shows the free energy and so on for this case. There exist four first-order 
phase transitions, as $\bar \beta$ is increased. At each time a transition undergoes, 
the number of non-zero $\psi^i$ increases by one, and the maximum number is four, which is the rank of $C$.

\begin{figure}
\begin{center}
\includegraphics[width=10cm]{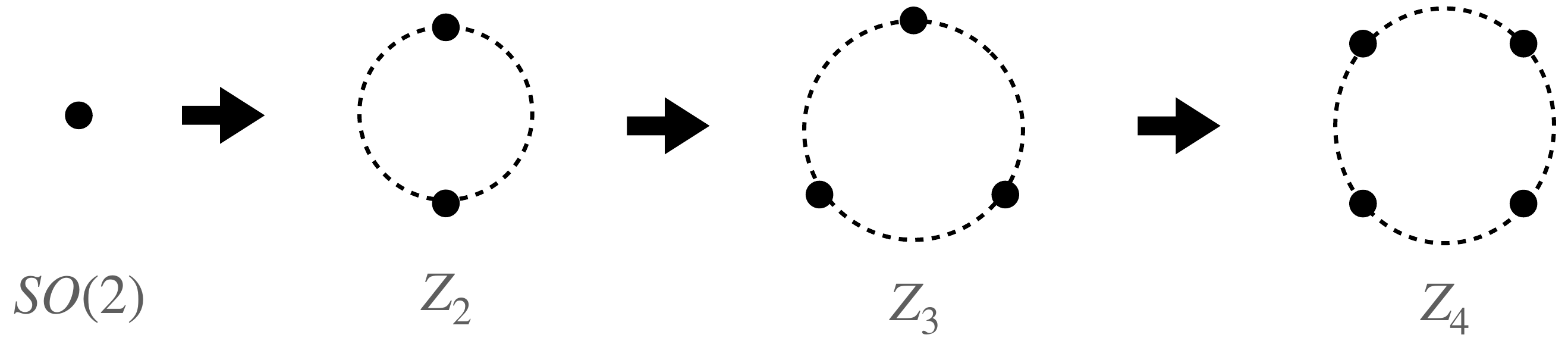}
\caption{Illustration of the development of the symmetry patterns in \eq{eq:symbreakso2}. Non-zero $\psi^i$'s 
form discretized $S^1$, the lattice spacing of which becomes finer as $\bar \beta$ is increased. The dots illustrate
the locations of non-zero $\psi^i$'s, and the dashed line an imaginary $S^1$. 
With increasing $\bar \beta$, an $S^1$ gradually emerges by the discrete steps of the first-order phase transitions.}
\label{fig:points}
\end{center}
\end{figure}

The symmetry breaking pattern of this case is more interesting than the previous cases. 
Because of the $SO(2)$ symmetry of $C$, the system initially has $SO(2)\times S_R$ symmetry. 
Since the non-zero $\psi^i$ in each phase are all different from each other, one would suspect that the fate of the 
real replica symmetry breaking would be the same as the previous cases. However, it is easy to check in each phase
that  the set of non-zero $\psi^i$ are invariant under a discrete subgroup of $SO(2)$.
For example, in the phase with two non-zero $\psi^i$s, say, $\psi^1,\psi^2$, they are related by 
$\psi^1=\hbox{Rot}(\pi)\psi^2$, where $\hbox{Rot}(\pi)$ is the $SO(2)$ rotation by angle $\pi$ (See Figure~\ref{fig:points}).
This is similar in the other phases, 
with three non-zero $\psi^i$'s being related by $\hbox{Rot}(\pi/3)$ and four non-zero $\psi^i$'s 
related by $\hbox{Rot}(\pi/4)$. In all, the symmetry breaking pattern is given by
\[
SO(2)\times S_R \rightarrow SO(2)\times S_{R-1} \rightarrow Z_2 \times S_{R-2} \rightarrow Z_3 \times S_{R-3}
\rightarrow Z_4 \times S_{R-4},
\label{eq:symbreakso2}
\]
where $Z_n$ denotes the cyclic group. As illustrated in Figure~\ref{fig:points}, 
one can check that non-zero $\psi^i$'s form discretized $S^1$, and the lattice spacing
becomes finer, as $\bar \beta$ is increased.
Note that, since there is a finite jump of configurations at each transition,
any direct connections (subgroup structre, etc.) do not generally exist between the consecutive breaking patterns.

\section{Random $C$}
\label{sec:randomC}
In this section we will consider the cases with random values of $C$. 
The motivation is to understand the dynamics of the system \eq{eq:system} for 
general values of $C$, rather than for some particular values as studied in Section~\ref{sec:excase1}.
Following the successes in the study of spin glasses, the main strategy we take is to apply the replica trick,
\[
\bar \beta F=-\left.\frac{\partial}{\partial T}\lim_{N\rightarrow \infty} \frac{1}{NR} \log \left \langle Z_{N,R}(\beta,C)^T \right\rangle_C \right|_{T=0},
\label{eq:replicatrick}
\]
where $\bar \beta=\beta/N$,
$\langle \cdot \rangle_C$ denotes the average over the random distribution of $C$, 
and $T$ is the replica number. As in \eq{eq:rangec},  we restrict the range of non-zero values of $C$ as 
$C_{abc}\ (a,b,c\leq n)$, and assume them to be distributed by the normal distribution:
\[
\langle {\cal O}\rangle_C = A \int \prod_{a\leq b \leq c=1}^n dC_{abc}\, {\cal O}\, e^{- \alpha C_{abc}C_{abc} },
\label{eq:distribution}
\]
where $\alpha$ is a positive number, and $A$ is a normalization factor for $\langle 1 \rangle_C=1$.

Let us first rewrite the power $T$ in \eq{eq:replicatrick} by introducing $T$ replicas of $\phi_a^i$:
\[
\left\langle Z_{N,R}(\beta,C)^T \right\rangle_C =A \int dC \, e^{-\alpha C^2}\, Z_{N,R}(\beta,C)^T =A \int dC \int_{I^T} \prod_{t=0}^{T-1}\prod_{a,i=1}^{N,R} d\phi^{it}_a \, e^{-S_T},
\]
where 
\s[
S_T=\alpha\,  C_{abc}C_{abc} +\beta \sum_{t=0}^{T-1}\left(C_{abc}-\sum_{i=1}^R \phi_a^{it} \phi_b^{it} \phi_c^{it}\right)^2,
\label{eq:start}
\s]
and we have introduced an additional upper index $t\ (t=0,1,\ldots,T-1)$ for $\phi$, which starts from zero
for later convenience.
With the same spirit as \eq{eq:division}, we divide $\phi$ into two parts,
\s[
&\psi_a^{it}=\phi_a^{it} \ (a=1,2,\ldots,n), \\
&\tilde \psi_a^{it}=\phi_a^{it}\ (a=n+1,n+2,\ldots,N).
\label{eq:divisiont}
\s]
After a straightforward computation, we obtain
\s[
S_T&=(\alpha +\beta T)\left( C_{abc}-\frac{\beta}{\alpha+\beta T} \sum_{i,t}\psi_a^{it}\psi_b^{it}\psi_b^{it}
\right)^2
-\frac{\beta^2}{\alpha+\beta T} \sum_{i,i',t,t'} \left(\psi^{it}\cdot \psi^{i't'} \right)^3 \\
&+\beta \sum_{i,i',t} \left( \psi^{it} \cdot \psi^{i't} \right)^3 
+3 \beta \sum_{i,i',t} \left( \psi^{it} \cdot \psi^{i't} \right)^2\left( \tilde \psi^{it} \cdot \tilde \psi^{i't} \right)\\
&+3 \beta \sum_{i,i',t} \left( \psi^{it} \cdot \psi^{i't} \right)\left( \tilde \psi^{it} \cdot \tilde \psi^{i't} \right)^2
+\beta \sum_{i,i',t} \left( \tilde \psi^{it} \cdot \tilde \psi^{i't} \right)^3,
\label{eq:sttenkai}
\s]
where the ranges of sums have been omitted for brevity, since they are obvious.

The first term of \eq{eq:sttenkai} can be integrated over $C$, and this generates 
\[
\frac{\#C}{2} \log\left(\frac{\alpha+\beta T}{\alpha}\right)
\]
as an additional term to the exponent, where $\#C=n(n+1)(n+2)/6$,  namely, the number of independent elements of $C$.
To compute the other terms, let us introduce the overlaps,
\s[
&Q_{iti't'}=\psi^{it}\cdot \psi^{i't'},\\
&\tilde Q_{iti't'}=\tilde \psi^{it}\cdot \tilde \psi^{i't'},
\label{eq:qpsi}
\s]
as in \eq{eq:tildeQ}. The same procedure as before generates similar logarithmic terms as in \eq{eq:seff},
\[
-\frac{n}{2} \log \det Q-\frac{N-n}{2}\log \det \tilde Q,
\label{eq:log}
\]
where $Q$ and $\tilde Q$ are regarded as $RT\times RT$ matrices in the determinants.
By assembling the above expressions, we obtain
\s[
S_T^{\rm eff}(\beta)=&\frac{\#C}{2} \log\left(1+\frac{\beta}{\alpha} T \right) 
-\frac{\beta^2}{\alpha+\beta T} \sum_{i,i',t,t'} \left(Q^{iti't'} \right)^3 
+\beta \sum_{i,i',t} \left(Q^{iti't} \right)^3 \\
&+3 \beta \sum_{i,i',t} \left(Q^{iti't} \right)^2\left( \tilde Q^{iti't} \right)
+3 \beta \sum_{i,i',t} \left( Q^{iti't} \right)\left( \tilde Q^{iti't} \right)^2
+\beta \sum_{i,i',t} \left( \tilde Q^{iti't} \right)^3 \\
&-\frac{n}{2} \log \det Q-\frac{N-n}{2}\log \det \tilde Q.
\label{eq:sefft1}
\s]

For later use, it is more convenient to perform a rescaling of variables in \eq{eq:sefft1}:
$Q=(N/\beta)^{1/3} \bar Q,\ \tilde Q=(N/\beta)^{1/3} \bar{\tilde Q}$. Then we obtain
\s[
S_T^{\rm eff}(\beta)=&\frac{\#C}{2} \log\left(1+\frac{\beta}{\alpha} T \right) +\frac{NRT}{6} \log \beta
-\frac{N \beta}{\alpha+\beta T} \sum_{i,i',t,t'} \left(\bar Q^{iti't'} \right)^3 
+N\sum_{i,i',t} \left(\bar Q^{iti't} \right)^3 \\
&+3 N \sum_{i,i',t} \left(\bar Q^{iti't} \right)^2\left( \bar{\tilde Q}^{iti't} \right)
+3N \sum_{i,i',t} \left( \bar Q^{iti't} \right)\left( \bar {\tilde Q}^{iti't} \right)^2
+N \sum_{i,i',t} \left(\bar{ \tilde Q}^{iti't} \right)^3 \\
&-\frac{n}{2} \log \det \bar Q-\frac{N-n}{2}\log \det \bar {\tilde Q},
\label{eq:sefft}
\s]
where we have ignored an unimportant constant shift. 

Further computations depend on assumptions made on replica symmetry breaking. Below
we consider only two possibilities, replica symmetric (RS) and one-step replica symmetry breaking (1RSB).
A reason to consider only these possibilities is that 1RSB is known to be exact \cite{pspin} in the spherical
$p$-spin model. In fact numerical simulations, which we will show in Section~\ref{sec:comparison},
seem to support this assumption at least in the large-$N$ limit
of this paper.

\subsection{$R=1$ with random $C$}

Since $R=1$, we can simply write $\bar Q^{tt'}$ in place of $\bar Q^{iti't'}$, neglecting the real replica index.  
Under the assumption of 1RSB \cite{pspin,pedestrians}, the overlap $\bar Q$ is assumed to have the form,
\[
\bar Q^{tt'}=\delta_{\lfloor t/M \rfloor, \lfloor t'/M \rfloor} I_{t \bmod  M, t' \bmod  M} (q_0,q_1)
+ q_2 (1-\delta_{\lfloor t/M \rfloor, \lfloor t'/M \rfloor}),
\label{eq:onestep} 
\]
where $q_i$ are new variables, $\lfloor \cdot \rfloor$ is the floor function, $a \bmod b$ denotes $a$ modulo $b$,
and $I$ is an $M\times M$ matrix with components,
\[
I_{tt'}(q_0,q_1)=q_0\, \delta_{tt'}+q_1(1-\delta_{tt'}).
\label{eq:matI}
\] 
Then it is straightforward to obtain 
\s[
&\frac{1}{T} \sum_{t,t'}(\bar Q^{tt'})^3=(T-M) q_2^3+(M-1) q_1^3+q_0^3 ,\\
&\frac{1}{T} \sum_{t} (\bar Q^{tt})^3=q_0^3, \\
&\log \det \bar Q=\log \left(q_0+(M-1) q_1+(T-M) q_2\right)+\left(\frac{T}{M} -1\right) \log \left( q_0-q_1+M(q_1-q_2)\right) \\
&\ \ \ \ \ \ \ \ \ \ \ \ \ \ \ \ +\left(T-\frac{T}{M}\right) \log (q_0-q_1).
\label{eq:qbarsum}
\s]
These expressions are also assumed for $\bar { \tilde Q}$, 
with replacements $q_i\rightarrow \tilde q_i, \ M\rightarrow \tilde M$.

By putting \eq{eq:qbarsum} and the corresponding expressions of $\bar {\tilde Q}$ to \eq{eq:sefft}, 
and applying the replica trick formula in \eq{eq:replicatrick}, one obtains
\s[
\bar \beta &F^{1RSB}=\frac{\# C}{2 N} \frac{\beta}{\alpha} +\frac{1}{6} \log \beta 
- \frac{\beta}{\alpha} \left( -M q_2^3+(M-1) q_1^3+q_0^3\right)
+q_0^3+3 q_0^2 \tilde q_0 +3 q_0 \tilde q_0^2+\tilde q_0^3 \\
&-\frac{\bar n}{2} \left( \left( 1-\frac{1}{M} \right) \log (q_0-q_1) +\frac{1}{M} \log ( 
q_0-q_1+M(q_1-q_2)) +\frac{q_2}{q_0+(M-1) q_1-M q_2}\right) \\
&-\frac{1-\bar n}{2} \left( \left( 1-\frac{1}{\tilde M} \right) \log (\tilde q_0-\tilde q_1) +\frac{1}{\tilde M} \log ( 
\tilde q_0-\tilde q_1+\tilde M(\tilde q_1-\tilde q_2)) +\frac{\tilde q_2}{\tilde q_0+(\tilde M-1) \tilde q_1-\tilde M \tilde q_2}\right),
\label{eq:freegeneral}
\s]
where we have introduced $\bar n=n/N$. 

Here let us comment on the ranges of the parameters which appear in the above expressions 
(See \cite{pedestrians} for more details). 
From the physical point of view, the overlaps of configurations in the same state will not be less than 
those in different states. Therefore
\[
q_0\geq q_1 \geq q_2,\ \tilde q_0\geq \tilde q_1 \geq \tilde q_2.
\label{eq:qrange}
\] 
In addition, the computations of
 $\langle \langle \psi_{a_1} \cdots \psi_{a_k} \rangle  \langle \psi_{a_1} \cdots \psi_{a_k} \rangle  \rangle_C$
 under the 1RSB assumption (and the same for $\tilde \psi$) 
 leads to a probability distribution of the overlap,
 \[
 \langle P(q) \rangle_C= \frac{M-1}{T-1} \delta (q-q_1) +\frac{T-M}{T-1} \delta(q-q_2),
 \] 
 and a similar one for $\tilde q_i$.
 Then the positivity of this probability distribution in the $T\rightarrow 0$ limit requires
 \[
 0\leq M \leq 1, \  0\leq \tilde M \leq 1.
 \label{eq:mrange}
 \]
One can check that \eq{eq:qrange} and \eq{eq:mrange} assures the arguments of the logarithms and 
the denominators in \eq{eq:freegeneral} do not become less than zero. 

For simplicity of discussions, in the rest of this paper, we assume $q_2,\tilde q_2=0$, since this is known to be 
true for the $p$-spin spherical model \cite{pspin}. The consistency of our analysis of this section and
the numerical results of Section~\ref{sec:comparison} will support this simplification.

When $\beta$ is small enough, namely, in high temperatures, the system will be in RS phase.
The RS expression of the free energy can be obtained by putting $q_1=\tilde q_1=0$ (and $q_2=\tilde q_2=0$)
to the 1RSB expression \eq{eq:freegeneral}:
\s[
\bar \beta F^{RS}=&\frac{\# C}{2 N} \frac{\beta}{\alpha} +\frac{1}{6} \log \beta +\left(1-\frac{\beta}{\alpha}\right)q_0^3
+3 q_0^2 \tilde q_0 +3 q_0 \tilde q_0^2+\tilde q_0^3 -\frac{\bar n}{2} \log q_0-\frac{1-\bar n}{2} \log \tilde q_0. 
\label{eq:freers}
\s]
Note that $M,\tilde M$ have disappeared from the expression, since they have no roles
for $q_1=\tilde q_1=q_2=\tilde q_2=0$ in \eq{eq:onestep}.

The free energy of the system is obtained by searching for the stationary points of these expressions 
\eq{eq:freegeneral} and \eq{eq:freers} in terms of the variables, $q_i,\tilde q_i,M,\tilde M$.

\subsubsection{$R=1$, $\bar n=1$ with random $C$}
\label{sec:r1nbar1}
In this subsection, let us perform concrete analysis 
on the case with $R=1,\ \bar n=1$ and random $C$ by using the expressions above.
Because of $\bar n=1$, we can simply ignore the terms with $\tilde q_i$.
Therefore, from \eq{eq:freegeneral} and \eq{eq:freers}, we have 
\[
\bar \beta F^{RS}_{R=1\, \bar n=1}=&\frac{\# C}{2 N} \frac{\beta}{\alpha} +\frac{1}{6} \log \beta +\left(1-\frac{\beta}{\alpha}\right)q_0^3
 -\frac{1}{2} \log q_0,
\label{eq:freersn1}
\]
and 
\s[
\bar \beta F^{1RSB}_{R=1\,\bar n=1}=&\frac{\# C}{2 N} \frac{\beta}{\alpha} +\frac{1}{6} \log \beta 
- \frac{\beta}{\alpha} \left((M-1) q_1^3+q_0^3\right)
+q_0^3 \\
&-\frac{1}{2M} \left( (M-1) \log (q_0-q_1) +\log ( 
q_0-q_1+M q_1) \right),
\label{eq:freern1}
\s]
where we have put $q_2=0$ following the spherical $p$-spin model \cite{pspin}. 

Let us first consider the RS case \eq{eq:freersn1}. By solving 
$\frac{\partial \bar \beta F_{R=1\,\bar n=1}^{RS}}{\partial q_0}=0$,
one obtains
\[
\bar \beta F^{RS}_{R=1\,\bar n=1}=\frac{\# C}{2 N} \frac{\beta}{\alpha} +\frac{1}{6} \log \beta+\frac{1}{6}
+\frac{1}{6} \log 6 + \frac{1}{6} \log \left(1-\frac{\beta}{\alpha}\right).
\] 
Because of the divergence at $\beta=\alpha$, the RS phase cannot exist for all $\beta$ and 
there must be a critical point $\beta_c<\alpha$, by which the RS phase is bounded.

A comment is in order. The appearance of this diverging point can be traced to the appearance of a curious 
indefinite coefficient $1-\beta/\alpha$ in \eq{eq:freersn1}. 
This looks confusing, because the starting expression \eq{eq:start} is positive semidefinite.
However, after integrating out $C$, we obtain \eq{eq:sefft1}, which contains a negative coefficient.
One can check that this expression is still positive semidefinite for $T\geq 1$, 
but after taking $T=0$ of the replica trick, it becomes 
indefinite. Therefore the presence of a phase transition 
is an interplay between integrating out and the analytic continuation.

We assume that the RS solution is connected to the 1RSB solution at $\beta=\beta_c<\alpha$ with the requirement that 
the free energy is continuous at this point.
The stationary condition of \eq{eq:freern1} is given by
\[
\frac{\partial \bar \beta F^{1RSB}_{R=1\,\bar n=1}}{\partial q_0}=\frac{\partial \bar \beta F^{1RSB}_{R=1\,\bar n=1}}{\partial q_1}=\frac{\partial \bar \beta F^{1RSB}_{R=1\,\bar n=1}}{\partial M}=0.
\label{eq:st1RSB}
\] 
These equations, however, cannot be fully solved algebraically, 
because of the logarithmic terms in the last equation. 
However, using Mathematica, the first two equations can analytically be solved, in which
$q_0,q_1$ are analytically expressed by $M$.
Then the last equation can be numerically solved to obtain $M$. With this procedure, one can 
compute the free energy for each value of $\alpha,\beta$.

\begin{figure}
\begin{center}
\includegraphics[width=5cm]{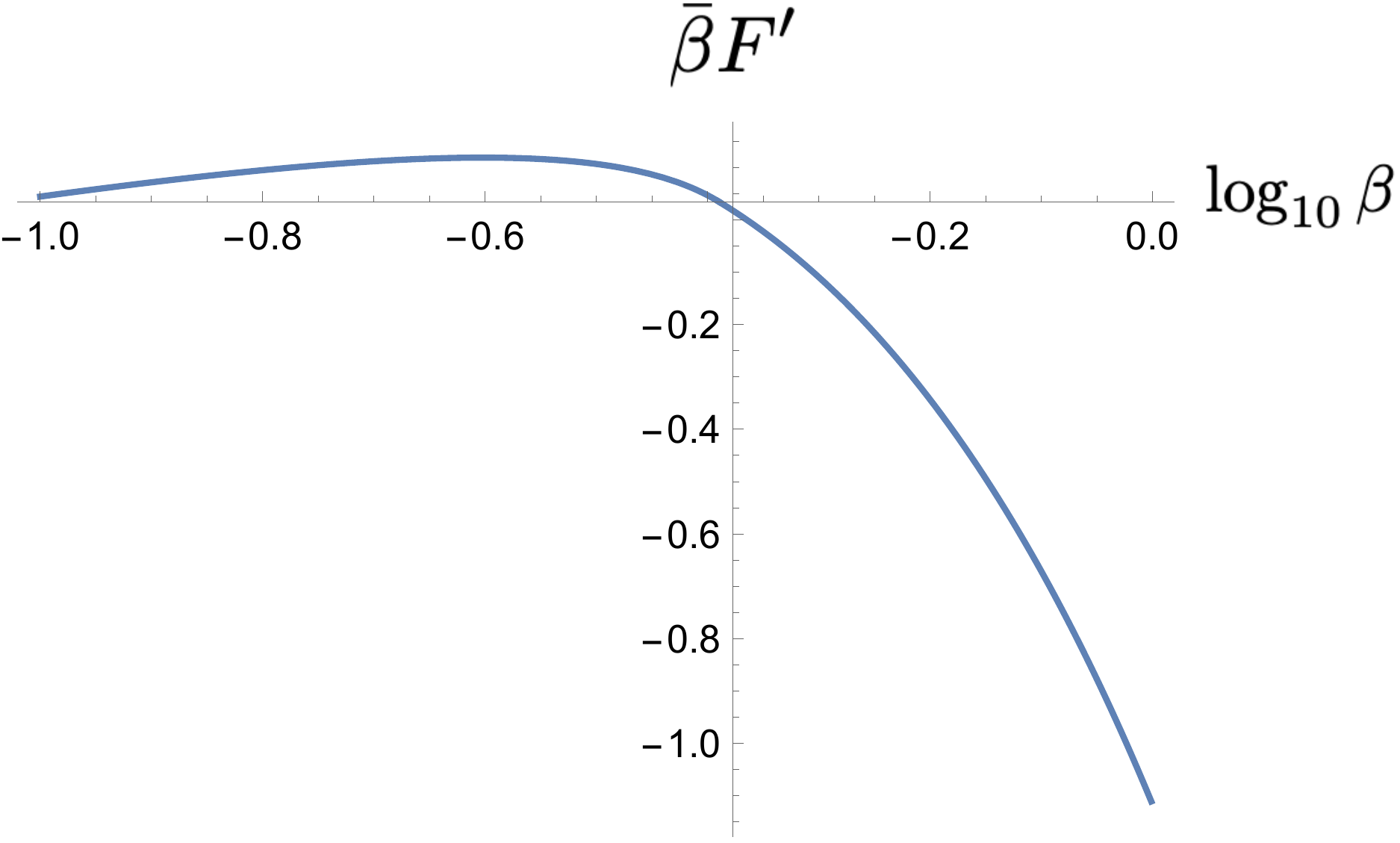}
\hfil
\includegraphics[width=5cm]{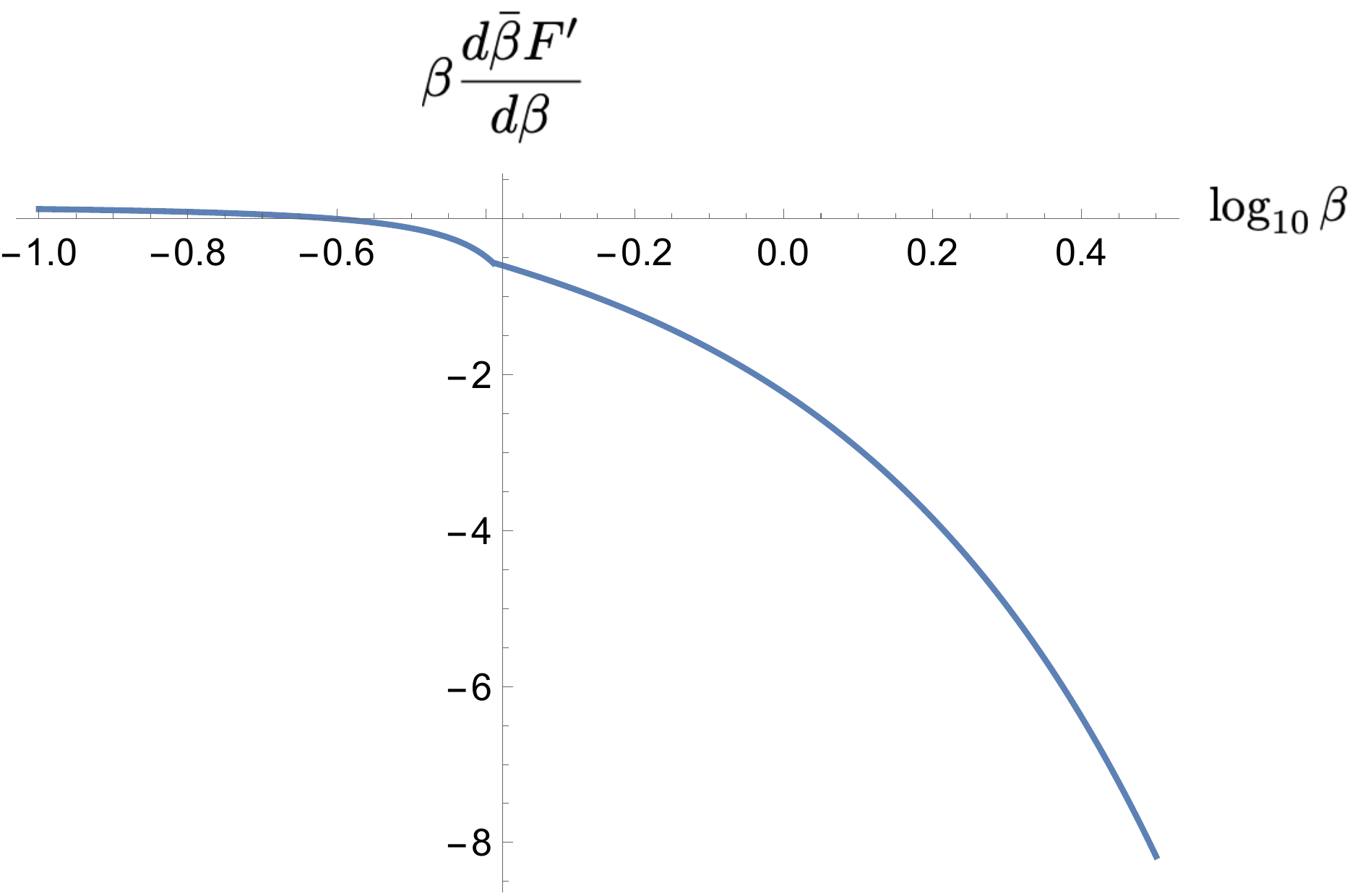}
\hfil
\includegraphics[width=5cm]{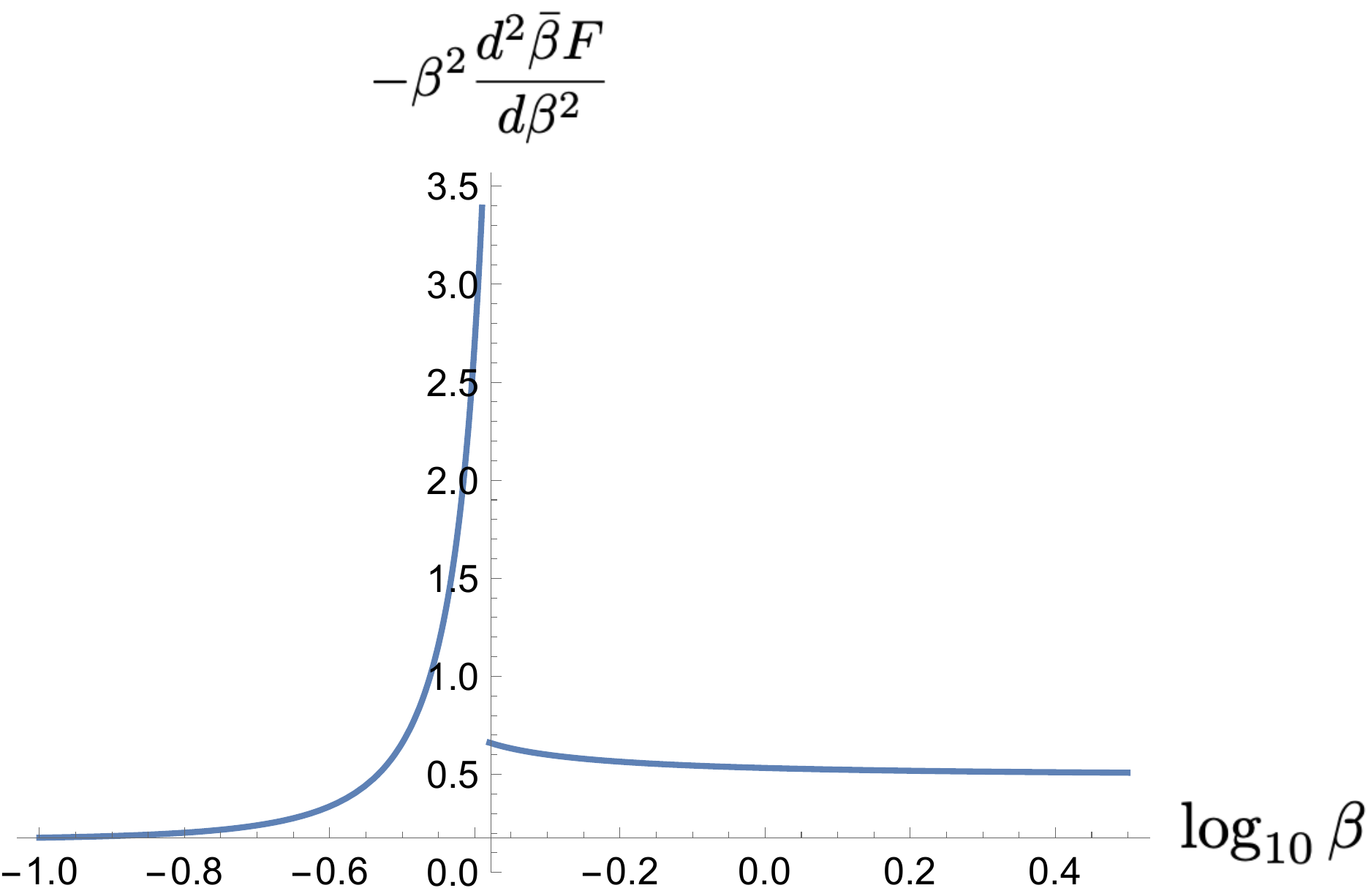}
\caption{The free energy and its first and second derivatives for  $R=1,\ \bar n=1$ and random $C$.
$\alpha=1/2$ is taken without loss of generality.
$F'$ denotes the free energy with the subtraction of the first term of \eq{eq:freersn1} and \eq{eq:freern1}. 
There is a second order phase transition with a finite jump
of the second derivative at $\beta_c\sim 0.407\ (\log_{10} \beta_c \sim -0.39)$. }
\label{fig:ran}
\end{center}
\end{figure}

\begin{figure}
\begin{center}
\includegraphics[width=7cm]{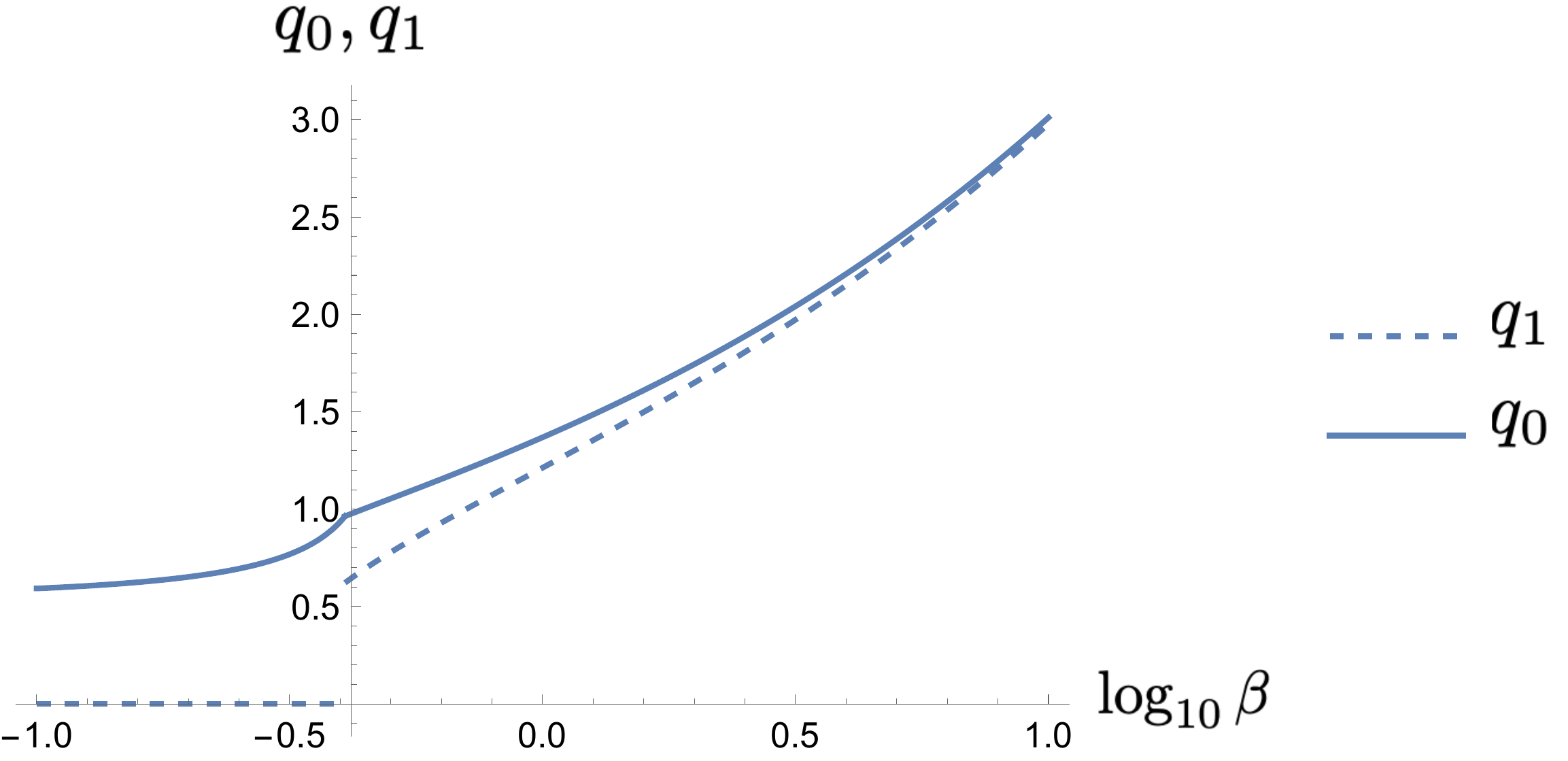}
\hfil
\includegraphics[width=7cm]{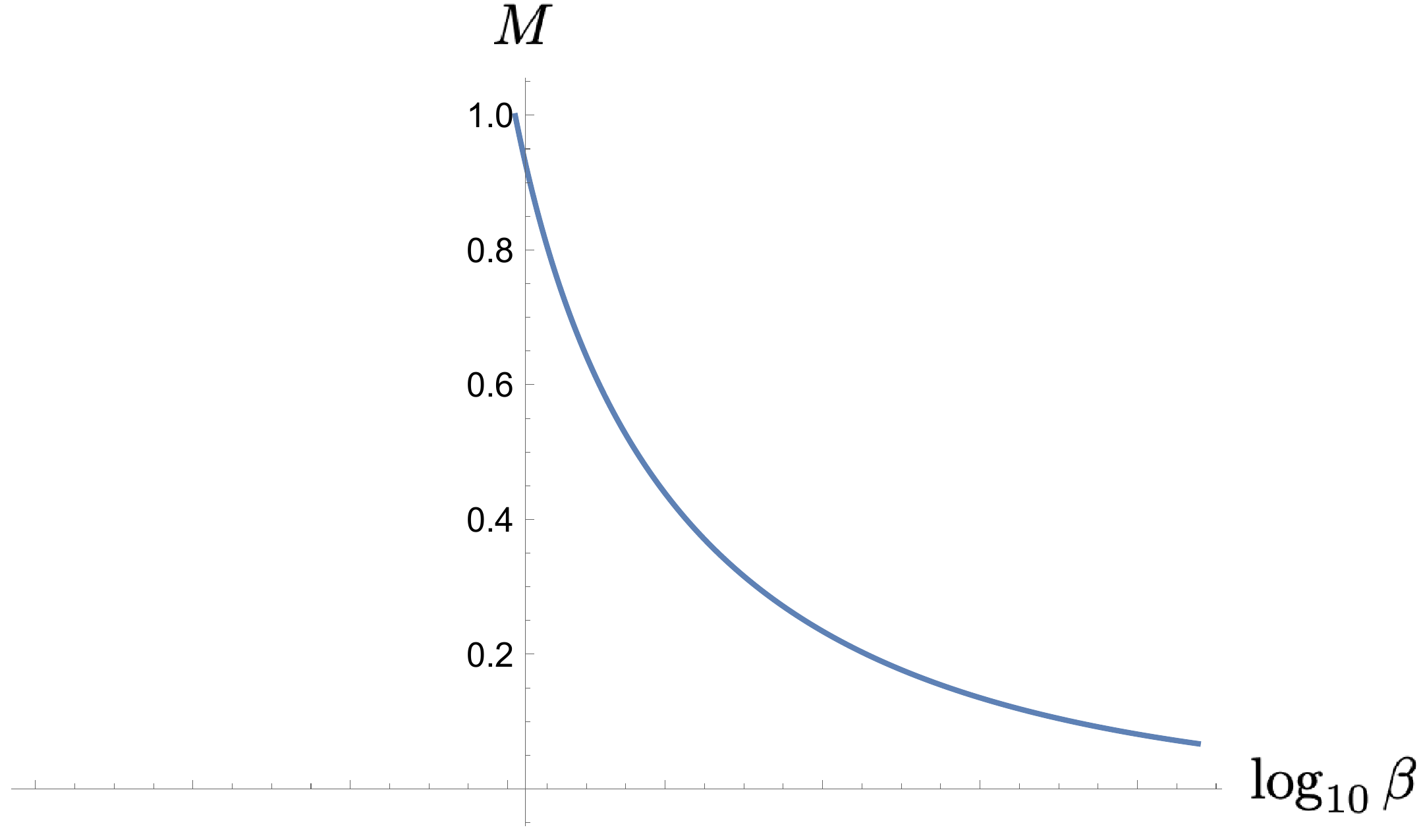}
\caption{The behavior of the configurations for the same case as in Figure~\ref{fig:ran}. $q_0$ is continuous,
while $q_1$ makes a jump at the transition point.}
\label{fig:ran2}
\end{center}
\end{figure}

Figure~\ref{fig:ran} shows the free energy up to the second derivatives. Here, we have subtracted the first term in
\eq{eq:freersn1} and \eq{eq:freern1} and have denoted the subtracted free energy by $F'$, 
because this first term behaves in $\sim N^2$ and is much larger than the other terms. 
Note that this term is linear in $\beta$ and therefore does not appear in the second derivative.  
To be concrete, we take $\alpha=1/2$ without loss of generality. 
The transition point can be determined by requiring the continuity of the free energy, $F^{RS}_{R=1\, \bar n=1}
=F^{1RSB}_{R=1\,\bar n=1}$, at $\beta=\beta_c$, and the result is $\beta_c\sim 0.407\ (\log_{10} 0.407 \sim -0.39)$.
The system is in the RS phase for $\beta<\beta_c$, and in the 1RSB phase for $\beta>\beta_c$.
The transition is second-order with a finite jump of the second derivative of the free energy with respect to $\beta$. 
The reason for the continuity of the first derivative of the free energy at the transition point can be attributed
to the continuity of $q_0$ and $M=1$, while
$q_1$ makes a jump, as shown in Figure~\ref{fig:ran2}.

\subsubsection{$R=1$, $\bar n=1/2$ with random $C$}
\label{sec:nhalf}
In this subsection, we consider a case with $R=1$, $\bar n<1$, and random $C$. To be concrete, we set $\bar n=1/2$ as 
a representative case of $\bar n < 1$.
The main difference from the $\bar n=1$ case is that we find a first-order phase transition
in place of the second-order one for $\bar n=1$.

Let us first reduce the general form of the free energy given in \eq{eq:freegeneral},
since it seems enough to assume $q_2,\tilde q_1,\tilde q_2=0$ for the present analysis.
Here the assumption $\tilde q_1,\tilde q_2=0$ of RS for the variable $\bar {\tilde Q}$ would be natural, 
because only its diagonal components appear in the interaction terms of \eq{eq:sefft}. 
By putting these assumptions and $\bar n=1/2$ to \eq{eq:freegeneral}, we obtain
\s[
\bar \beta &F^{1RSB}_{R=1\, \bar n=1/2}=\frac{\# C}{2 N} \frac{\beta}{\alpha} +\frac{1}{6} \log \beta 
- \frac{\beta}{\alpha} \left((M-1) q_1^3+q_0^3\right)
+q_0^3+3 q_0^2 \tilde q_0 +3 q_0 \tilde q_0^2+\tilde q_0^3 \\
&-\frac{1}{4M} \left( \left( M-1 \right) \log (q_0-q_1) +\log ( 
q_0-q_1+Mq_1) \right)-\frac{1}{4}  \log \tilde q_0.
\label{eq:nhalffree}
\s]
The RS case can simply be obtained by putting $q_1=0$ to \eq{eq:nhalffree}:
\s[
\bar \beta &F^{RS}_{R=1\,\bar n=1/2}
=\frac{\# C}{2 N} \frac{\beta}{\alpha} +\frac{1}{6} \log \beta 
+\left(1- \frac{\beta}{\alpha}\right)q_0^3
+3 q_0^2 \tilde q_0 +3 q_0 \tilde q_0^2+\tilde q_0^3 
-\frac{1}{4}\log q_0-\frac{1}{4}  \log \tilde q_0.
\label{eq:nhalffreers}
\s]

In the RS case, the equations, $\partial \bar \beta F^{RS}_{R=1\,\bar n=1/2}/\partial q_0
=\partial \bar \beta F^{RS}_{R=1\,\bar n=1/2}/\partial \tilde q_0
=0$, can be solved algebraically for $q_0,\tilde q_0$, and the free energy can be obtained as a function of $\beta/\alpha$,
though the expression is too complicated to explicitly write down here. 

\begin{figure}
\begin{center}
\includegraphics[width=5cm]{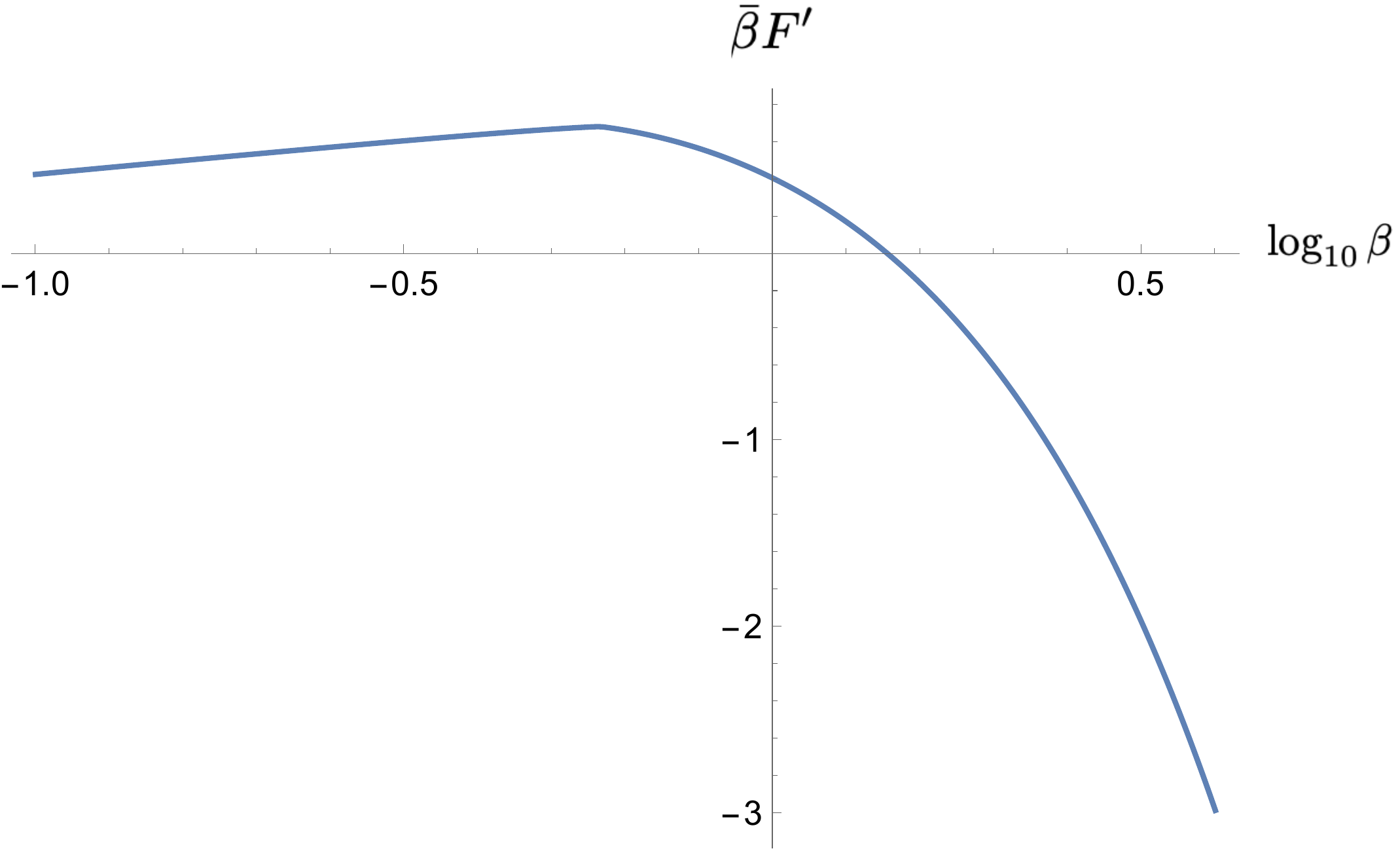}
\hfil
\includegraphics[width=5cm]{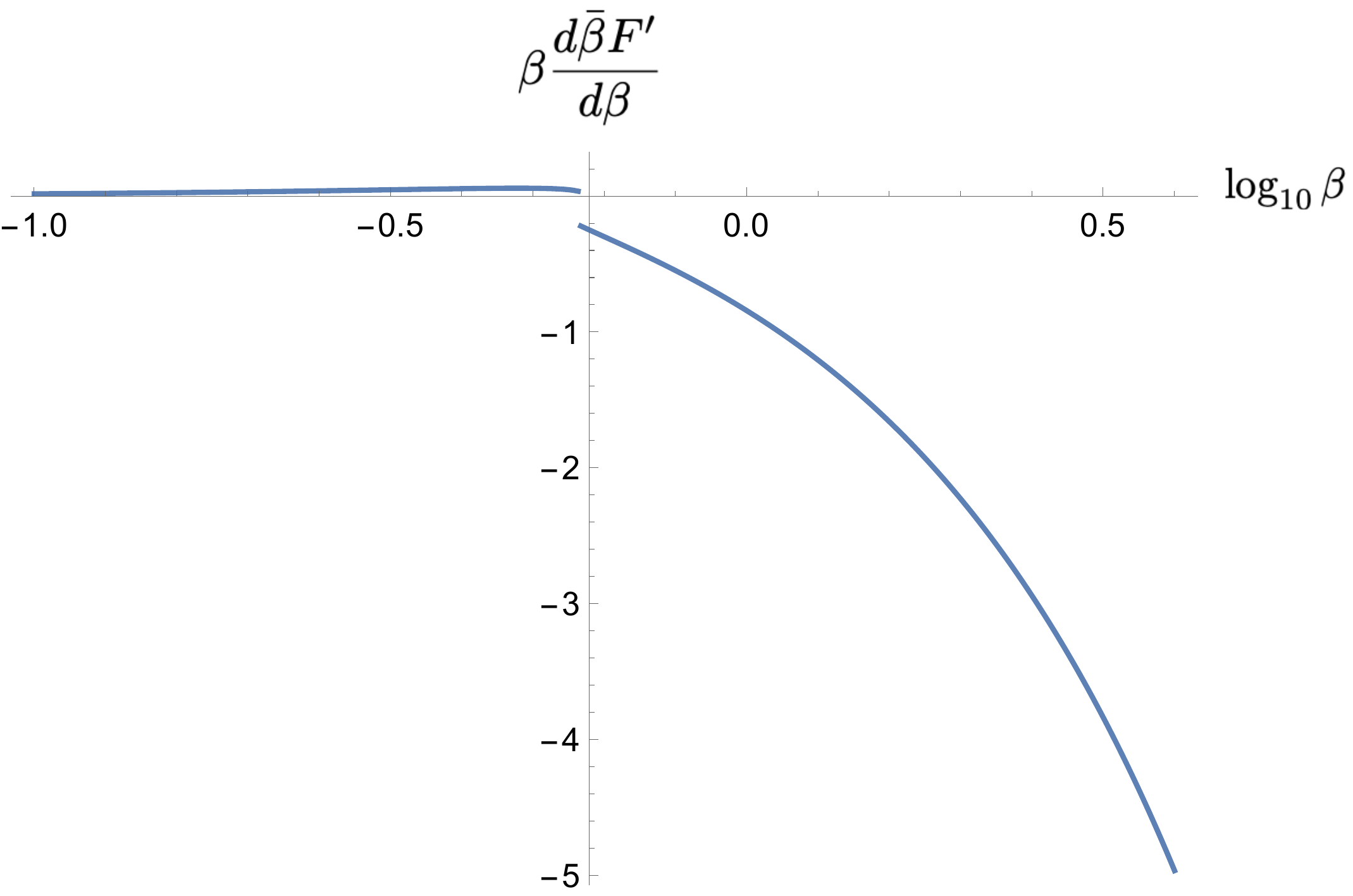}
\hfil
\includegraphics[width=5cm]{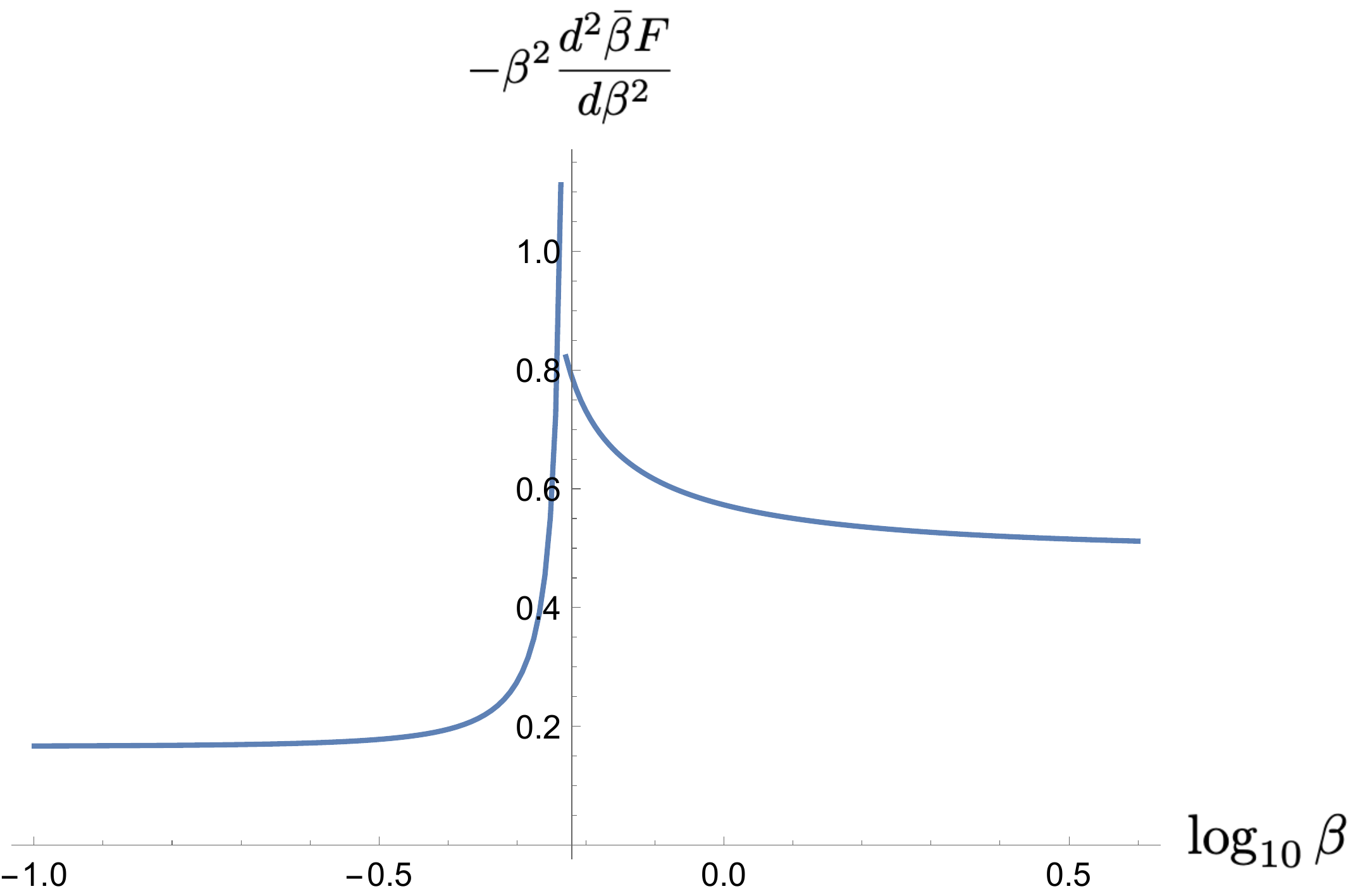}
\caption{The free energy and its first and second derivatives for $R=1,\ \bar n=1/2$. 
Without loss of generality, $\alpha=1/2$ is taken to be concrete.
A first-order phase transition exists at $\beta=\beta_c\sim  0.58\ (\log_{10} \beta_c \sim -0.24)$.
Here $F'$ denotes the free energy with the subtraction of the first term. }
\label{fig:nhalffree}
\end{center}
\end{figure}

\begin{figure}
\begin{center}
\includegraphics[width=5cm]{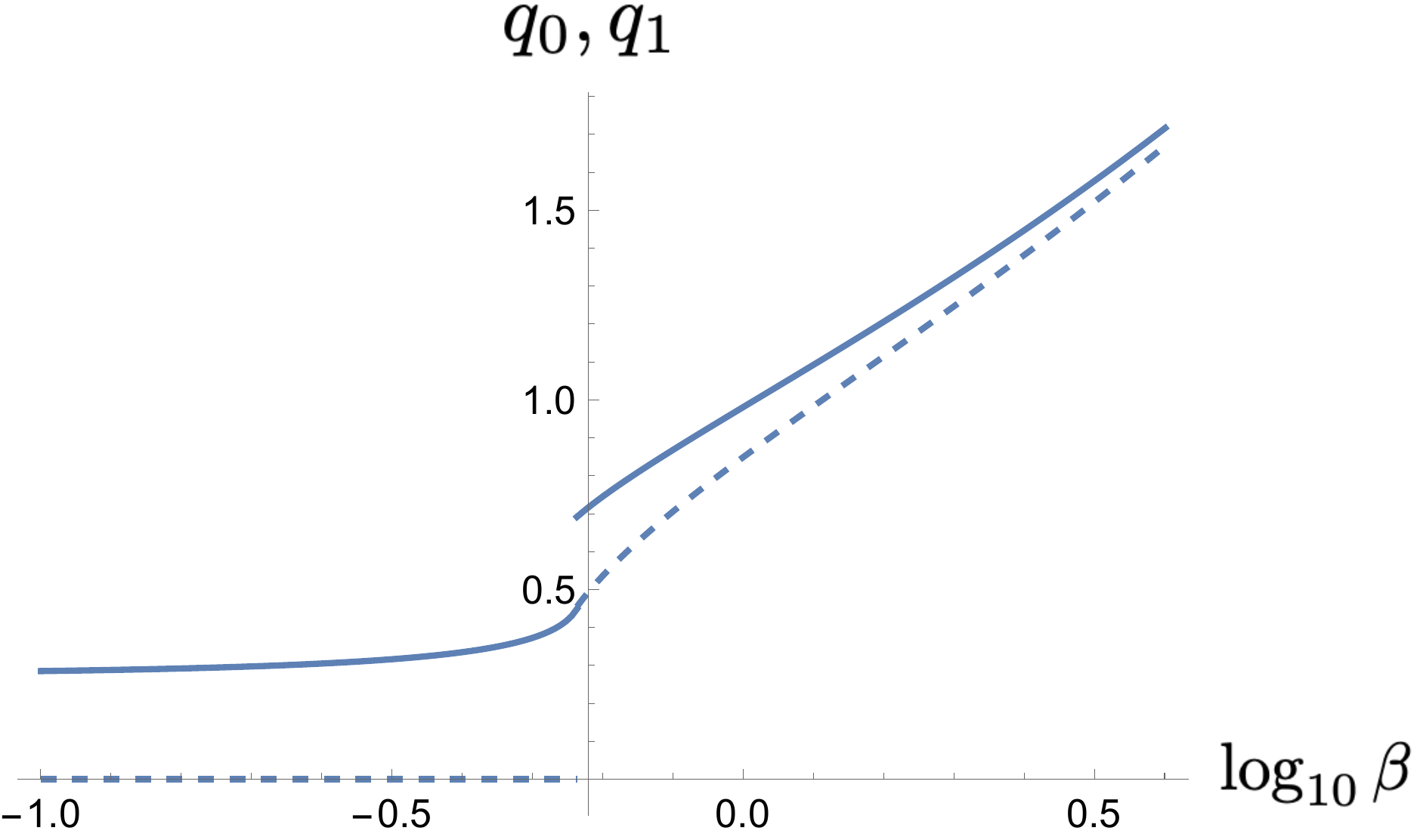}
\hfil
\includegraphics[width=5cm]{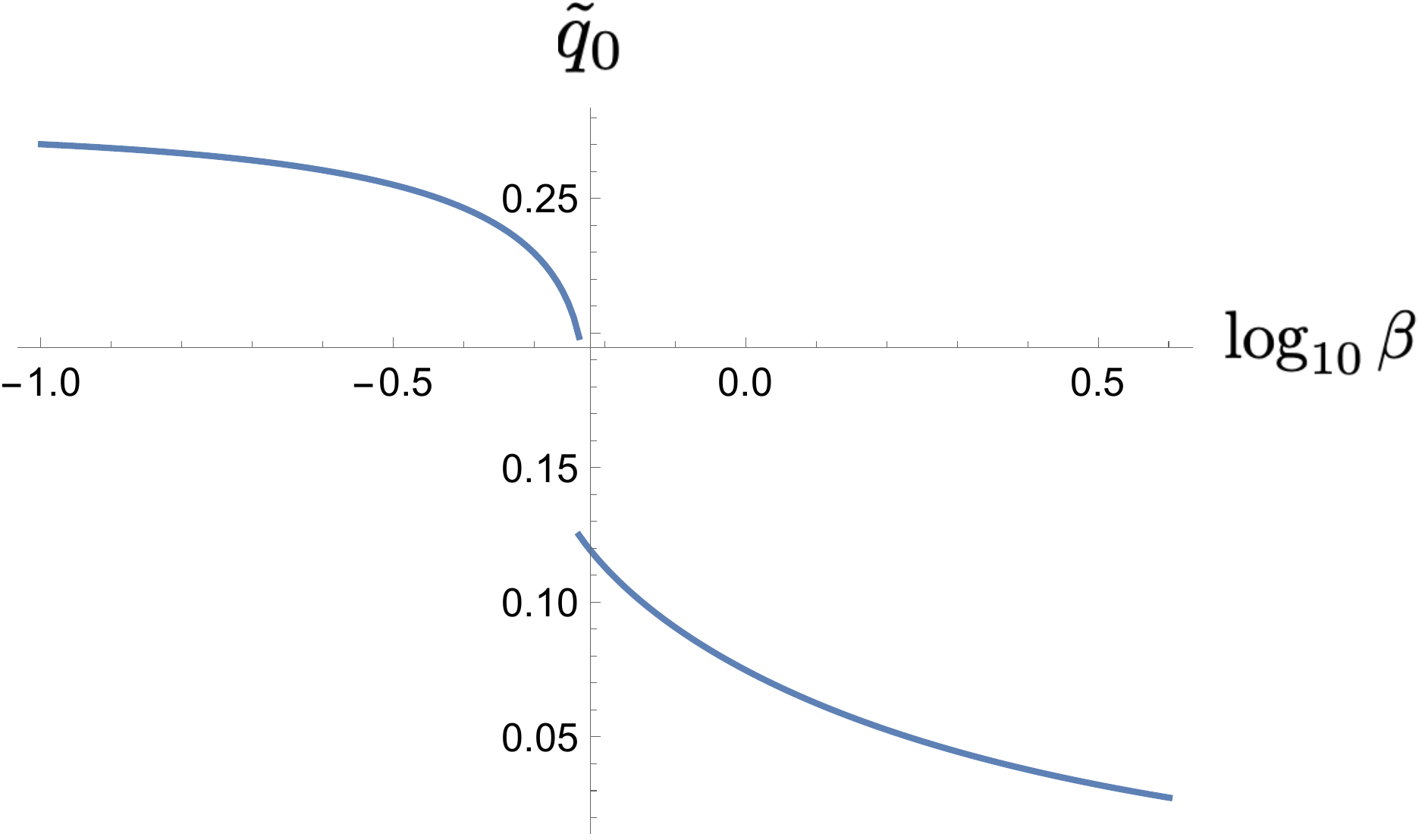}
\hfil
\includegraphics[width=5cm]{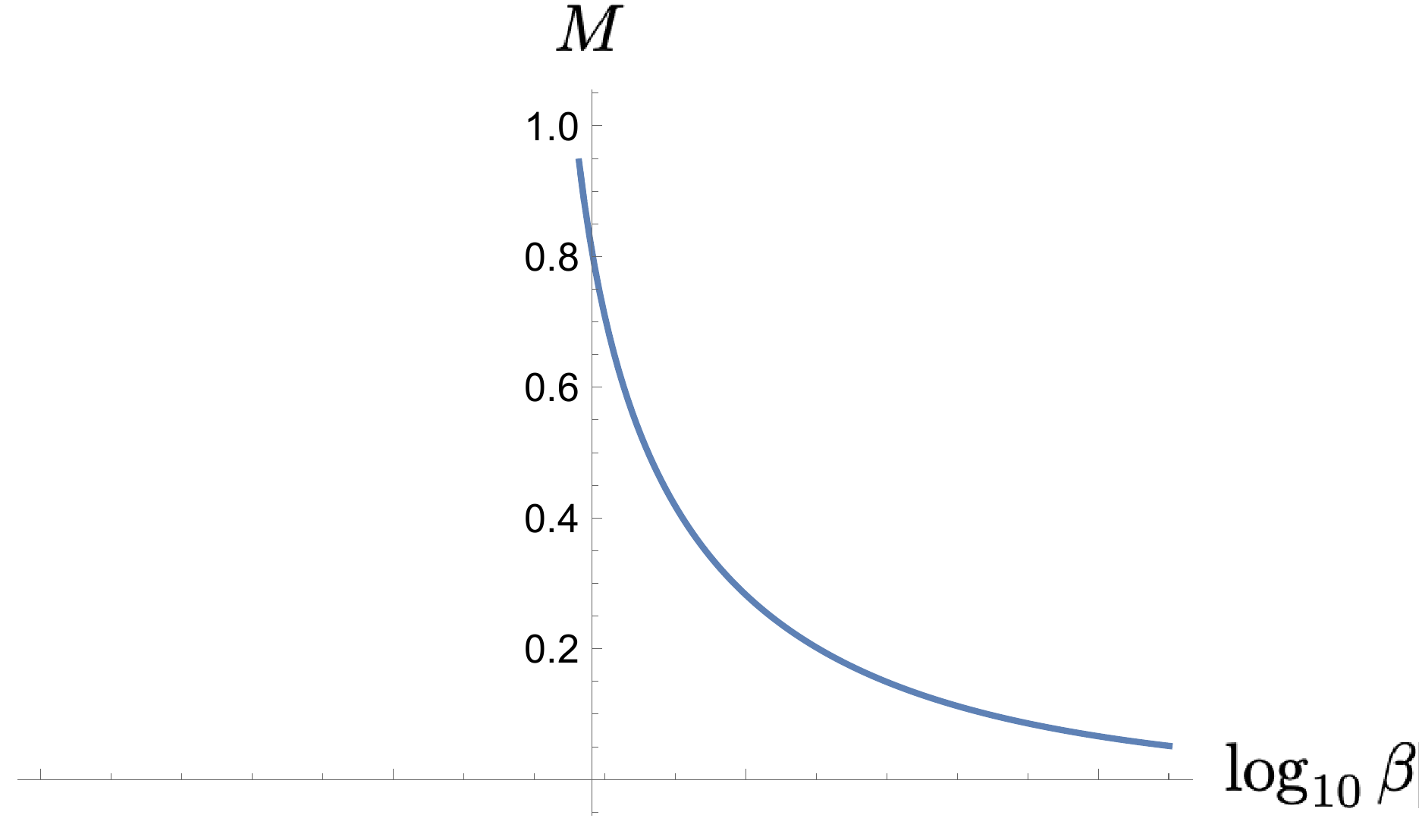}
\caption{The configurations are plotted for the same case as in Figure~\ref{fig:nhalffree}. 
$q_0$ and $q_1$ are shown by solid and dashed lines, respectively, in the left panel. 
The middle and right panels show $\tilde q_0$ and $M$, respectively.
The configurations jump at the phase transition point.
 }
\label{fig:nhalfconf}
\end{center}
\end{figure}

As for the 1RSB case, we have to fully resort to numerical analysis. 
To be concrete, let us take $\alpha=1/2$ without loss of generality.
Then we can find that there is a phase transition between RS and 1RSB phases at $\beta=\beta_c\sim  0.58\ (\log_{10}
\beta_c \sim -0.24)$,
as is shown in Figure~\ref{fig:nhalffree}. In the figure, the first terms of \eq{eq:nhalffree} and \eq{eq:nhalffreers} 
are subtracted in showing the free energy and its first derivative. The transition is first-order, as can be 
clearly seen in the plots (see also Figure~\ref{fig:nhalfconf}). 

\subsection{$R>1,\ \bar n=1$ with random $C$}
\label{sec:rlargenbar1}
To analyze the $R>1,\ \bar n=1$ case we assume the following form for $\bar Q$:
\s[
\bar{Q}^{iti't'}&=\delta_{\lfloor t/M \rfloor, \lfloor t'/M \rfloor} \left( \delta_{ii'} I_{t \bmod  M, t' \bmod  M} (q_0,q_1)+
(1-\delta_{ii'}) I_{t \bmod  M, t' \bmod  M} (q_0',q_1') \right),
\label{eq:qform}
\s] 
were the matrix $I$ is defined in \eq{eq:matI}, and 
we have already assumed that the $q_2,q_2'$ terms are not needed as in the previous cases. 
Here note that we are assuming that the real replica symmetry concerning the $i,i'$ indices of $\bar Q^{ii'tt'}$ 
is not broken, while the replica symmetry can be broken in general concerning $t,t'$.

Putting some formula shown in \ref{app:der} to \eq{eq:sefft} and using \eq{eq:replicatrick},
we obtain the free energy as
\s[
\bar \beta F_{R>1, \bar n=1}^{1RSB}&=\frac{\# C}{2 N R} \frac{\beta}{\alpha}+\frac{1}{6} \log \beta- \frac{\beta}{\alpha} \left((M-1) q_1^3+q_0^3\right)+q_0^3\\
&+(R-1)\left( 
- \frac{\beta}{\alpha} \left((M-1) q'_1{}^3+q'_0{}^3\right)+q'_0{}^3 \right)\\
&-\frac{1}{2MR}\Big[
\log\left\{q_0+(R-1)q_0'+(M-1)(q_1+(R-1)q_1')\right\}  \\
&\ \ \ \ \ \ \ \ \ \ +(M-1) \log \left\{q_0+(R-1)q_0'-(q_1+(R-1)q_1')\right\} \\
&\ \ \ \ \ \ \ \ \ \ +(R-1) \log \left\{q_0-q_0'+(M-1)(q_1-q_1')\right\}\\
&\ \ \ \ \ \ \ \ \ \ +(M-1)(R-1) \log \left\{ q_0-q_0'-(q_1-q_1') \right\}\Big].
\label{eq:freerlarge}
\s]
The RS case can be obtained by putting $q_1=q_1'=0$ as
\s[
\bar \beta F_{R>1, \bar n=1}^{RS}&=\frac{\# C}{2 N R} \frac{\beta}{\alpha}
+\frac{1}{6} \log \beta+\left(1- \frac{\beta}{\alpha} \right)q_0^3
+(R-1)\left(1- \frac{\beta}{\alpha} \right)q'_0{}^3 \\
&-\frac{1}{2R}\Big[
\log\left(q_0+(R-1)q_0'\right) +(R-1) \log \left(q_0-q_0'\right) \Big].
\label{eq:freerlarge0}
\s] 

In a similar manner as in the previous sections, one can numerically analyze the stationary solutions of 
\eq{eq:freerlarge} and \eq{eq:freerlarge0}. Then what we find is that there are only solutions with $q_0'=q_1'=0$. 
Therefore the present case reduces to the case with $R=1,\ \bar n=1$ case in Section~\ref{sec:r1nbar1}, 
as \eq{eq:freerlarge} and \eq{eq:freerlarge0} reduce to \eq{eq:freern1} and \eq{eq:freersn1}, respectively. 

\subsection{$R>1,\ \bar n<1$}
\label{sec:rlargensmall}
In this case, we have $\bar{\tilde Q}$ in addition.
Since only the diagonal components of $\bar{\tilde Q}$ with respect to $tt'$ indices couple with $\bar Q$ in \eq{eq:sefft},
it would be reasonable to assume the following diagonal form for $\bar {\tilde Q}$:
\[
\bar {\tilde Q}^{iti't'}=\delta_{tt'}\left( q_0 \delta_{ii'}+(1-\delta_{ii'}) q_0' \right).
\]
Then the interaction terms are given by
\s[
&3  \sum_{i,i',t} \left(\bar Q^{iti't} \right)^2\left( \bar{\tilde Q}^{iti't} \right)
+3 \sum_{i,i',t} \left( \bar Q^{iti't} \right)\left( \bar {\tilde Q}^{iti't} \right)^2
+ \sum_{i,i',t} \left(\bar{ \tilde Q}^{iti't} \right)^3  \\
&=RT\left( 3\left(q_0^2\tilde q_0+(R-1){q'_0}^2\tilde q'_0\right)+3\left(q_0\tilde q_0^2+(R-1)q'_0\tilde q'_0{}^2\right)+
\left(\tilde q_0^3+(R-1)\tilde q'_0{}^3\right)\right),
\label{eq:nbarint}
\s] 
and the determinant term of $\bar {\tilde Q}$ is given by 
\[
\log \det \bar {\tilde Q}=T \log (\tilde q_0+(R-1) \tilde q_0')+(R-1) T \log (\tilde q_0-\tilde q_0').
\label{eq:detrlargenbar}
\]
By appropriately including the parameter $\bar n$, the expression of the free energy can straightforwardly 
be obtained from \eq{eq:freerlarge}, \eq{eq:nbarint} and \eq{eq:detrlargenbar} in the same manner as before.
Then we numerically searched for the stationary solutions. However, all the solutions have 
$q_0'=q_1'=\tilde q_0'=0$. This concludes that the present case also reduces to the $R=1$ case.  

\section{Comparisons with numerical simulations}
\label{sec:comparison}
In this section, we perform some numerical simulations, and compare with the exact results.
By this we can check the exact results obtained based on some assumptions in the previous sections,
and can also check the large-$N$ convergence explicitly. 
The latter will be useful for future study, since this provides us some information on 
how much we can rely on numerical simulations in the analysis of our system.

We apply Hamiltonian Monte Carlo method \cite{Neal(2011)}  
to the system \eq{eq:system}, accompanied with parallel tempering \cite{paralleltemp}
across different values of $\beta$. We compute $\beta \langle s \rangle$ and 
$\beta^2 \langle s^2 \rangle_c=\beta^2(\langle s^2 \rangle-\langle s \rangle^2)$ with
$s=\beta (Q-\phi\phi\phi)^2$ to compare with
$\beta\frac{ \partial \beta F}{\partial \beta},\ -\beta^2 \frac{\partial^2 \beta F}{\partial \beta^2}$
of the exact results, respectively, where $\langle \cdot \rangle$ denotes the statistical average in the Monte Carlo
simulations. We also compute the $\beta$ dependencies of the configurations. 
The error estimates are performed by the Jackknife method.

As for random $C$, every independent component 
of $C$ is generated by the normal distribution with mean zero and standard deviation $1/\sqrt{m}$,
where $m$ is the multiplicity of the components, namely, 
$m(C_{iii})=1,m(C_{iij})=3, m(C_{ijk})=6$ for different $i,j,k$\footnote{This rescaling is necessary, because the $SO(N)$ invariant expression has the form, 
$C_{abc}C_{abc}=C_{iii}{}^2+3\, C_{iij}{}^2+6\,C_{ijk}{}^2$ due to $C$ being a symmetric tensor.}.
This distribution corresponds to taking $\alpha=1/2$, which was indeed taken in the plots shown in the previous 
sections. In general the details of the results depend on each particular randomly generated $C$, 
but we find that, when $N$ is large enough, such sample dependence seems irrelevant at least for the data we will show,
except for one quantity in Section~\ref{sec:numnbar1rlarge}.
We will actually see more or less good agreement with the exact results, even if we take one sample, 
except for the quantity in Section~\ref{sec:numnbar1rlarge}.

The machine had a Xeon W2295 (3.0GHz, 18 cores), 128GB DDR4 memory, and Ubuntu 20 as OS. 
The program was written in C++ with the use of pthread for parallelization.  
The leapfrog numbers were taken around a hundred. The numbers of the samples of each run are $10^4$-$10^6$.  
Every run typically took for 3-20 hours\footnote{The longest run was about 50 hours, which was for $N=70,\ R=2$.} 
with active use of parallelization, without serious tunings of the speed of the program.
 
We consider some representative cases in the following subsections.

\subsection{$n=2, \ R=4$, fixed $C$}  
\label{sec:n2r4sim}

This case corresponds to the exact results in Section~\ref{sec:neq2fixedc}. 

\begin{figure}
\begin{center}
\includegraphics[width=7cm]{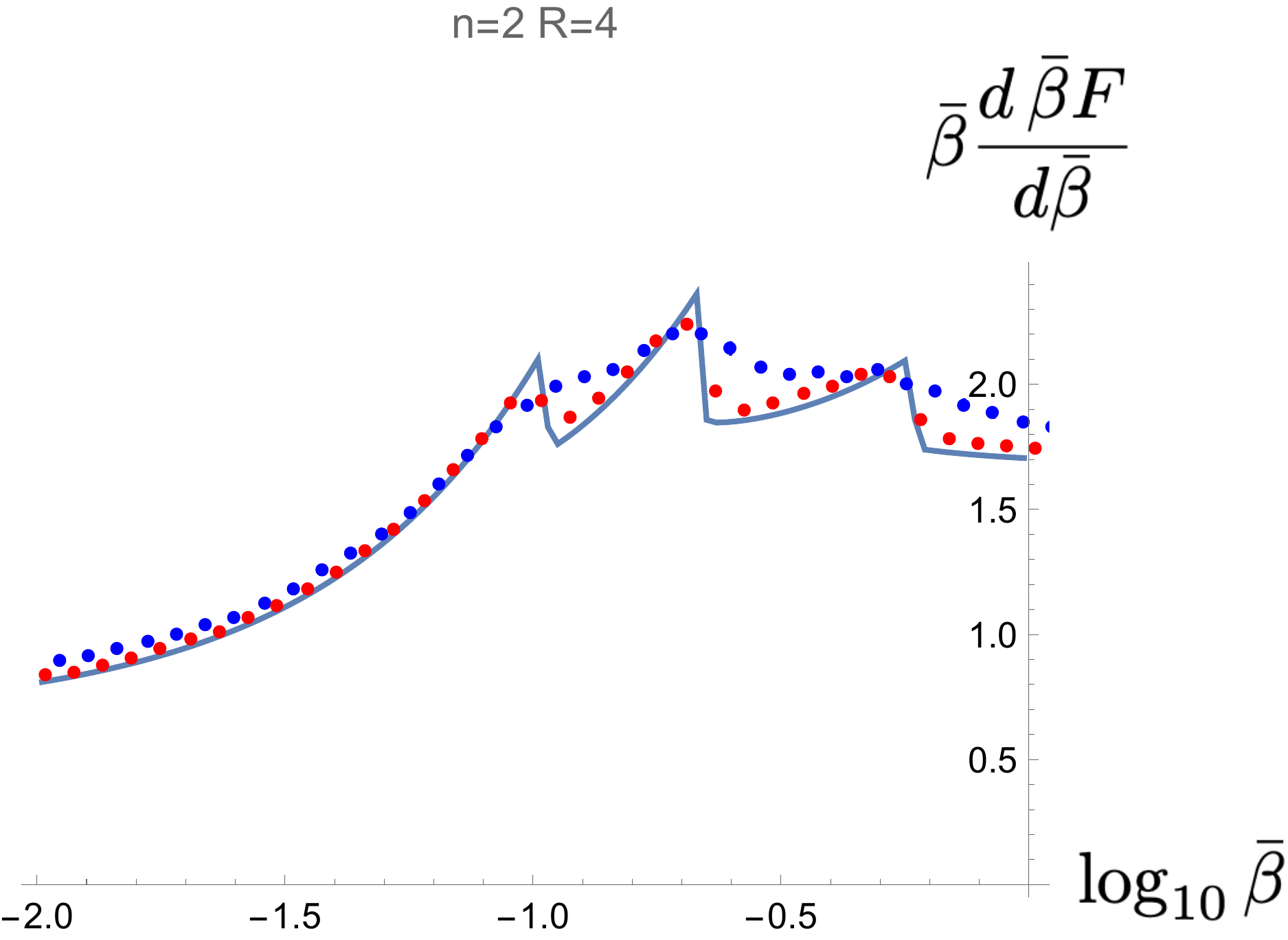}
\hfil
\includegraphics[width=7cm]{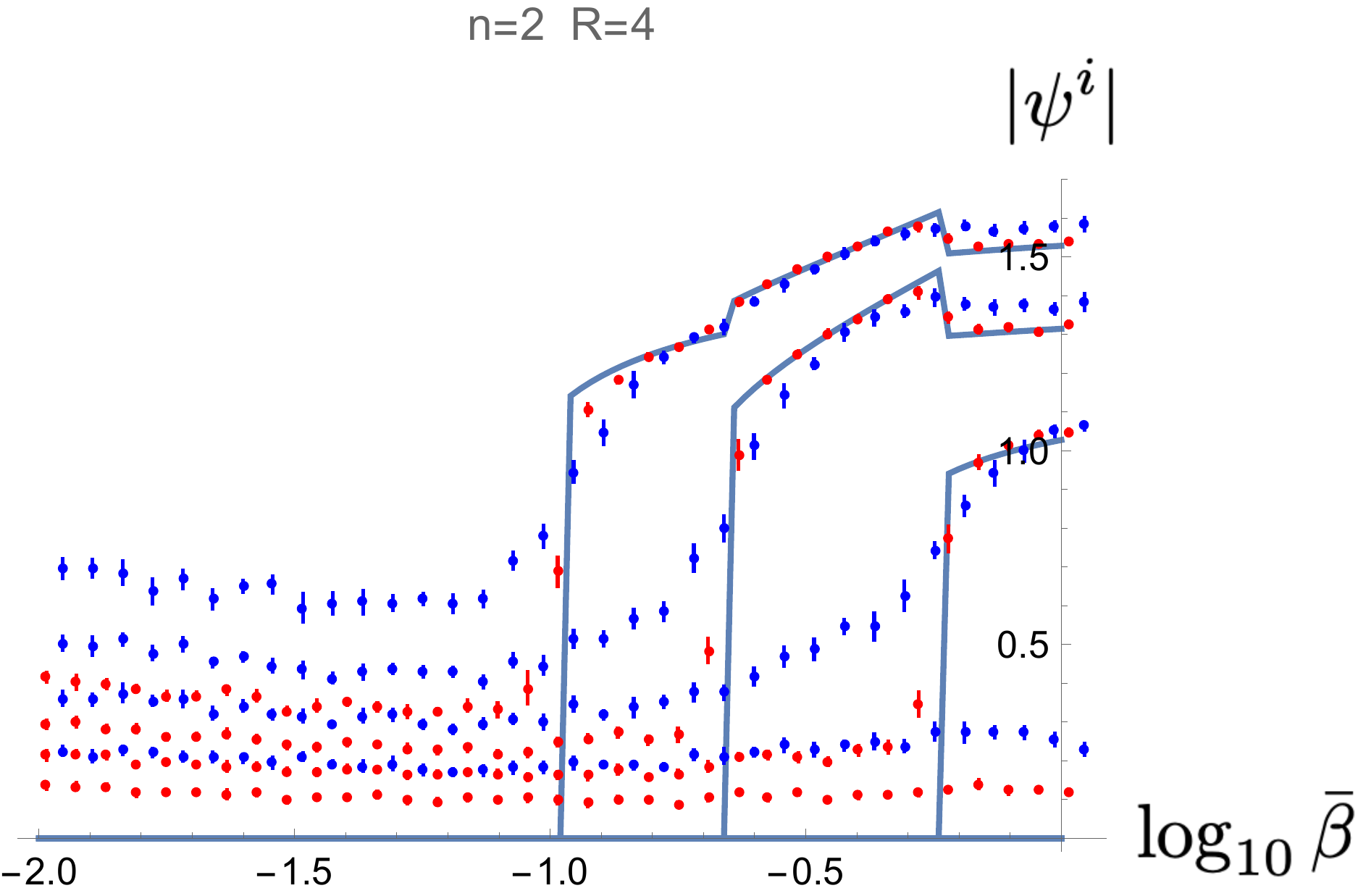}
\caption{The case with $n=2,\ R=4$ with fixed $C$. 
Comparisons between the exact results of Section~\ref{sec:neq2fixedc} and the numerical simulations
shown by dots with error bars. $N=20$ results are shown blue, and $N=60$ shown red. 
 }
\label{fig:n2}
\end{center}
\end{figure}

Figure~\ref{fig:n2} shows the comparison with the numerical 
simulations. The convergence of the free energy (first derivative) to the exact result is rather slow (the left panel), 
and it would be difficult to conclude the presence of multiple
first-order transitions solely from the data. On the other hand, from the plot of the configurations (the right panel), 
one would better be able to conclude there are jumps, implying first-order phase transitions.

\subsection{$\bar n=1,\ R=1$, random $C$}
\label{sec:neq1data}
This corresponds to the exact results in Section~\ref{sec:r1nbar1}. 

\begin{figure}
\begin{center}
\includegraphics[width=5cm]{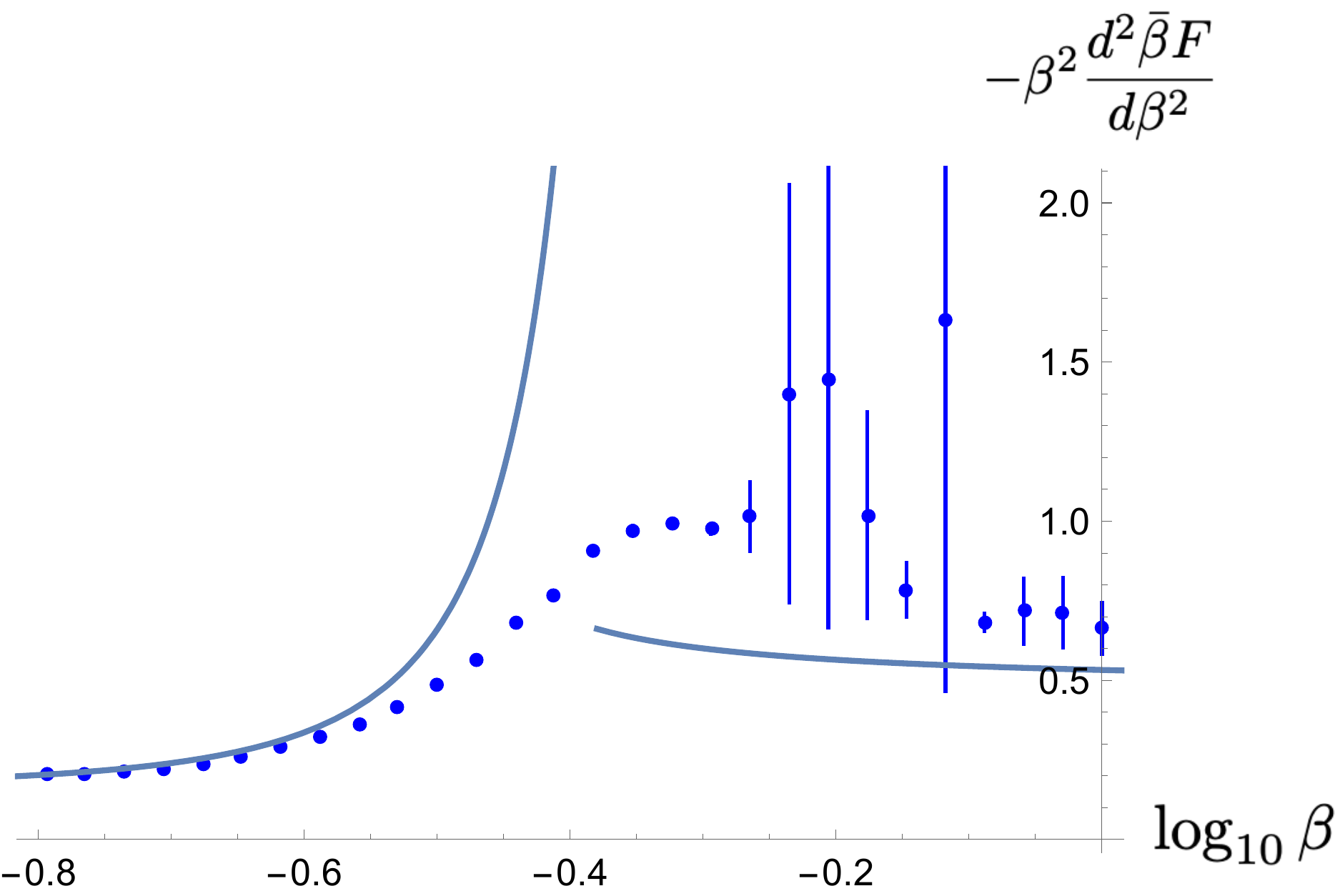}
\hfil
\includegraphics[width=5cm]{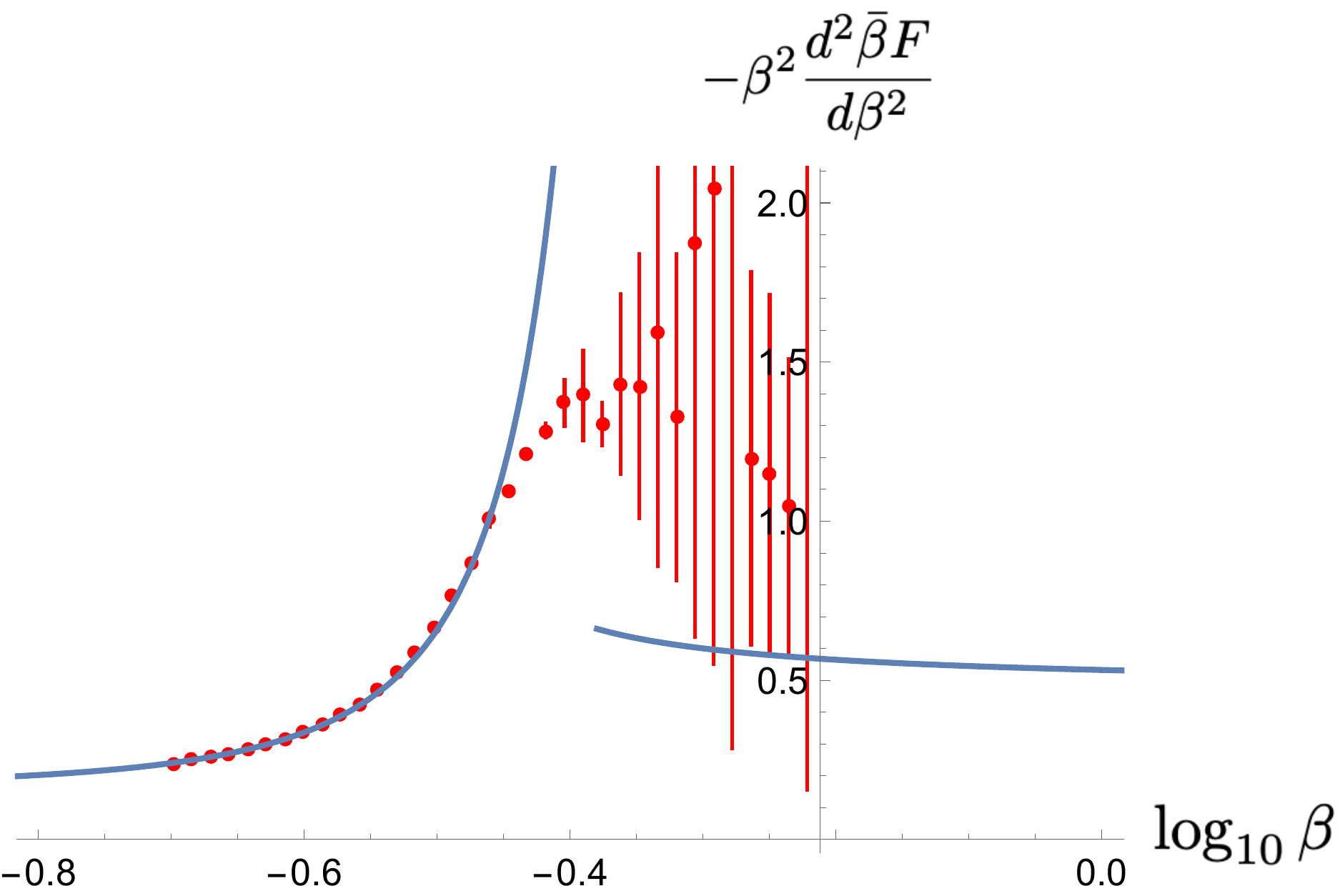}
\hfil
\includegraphics[width=5cm]{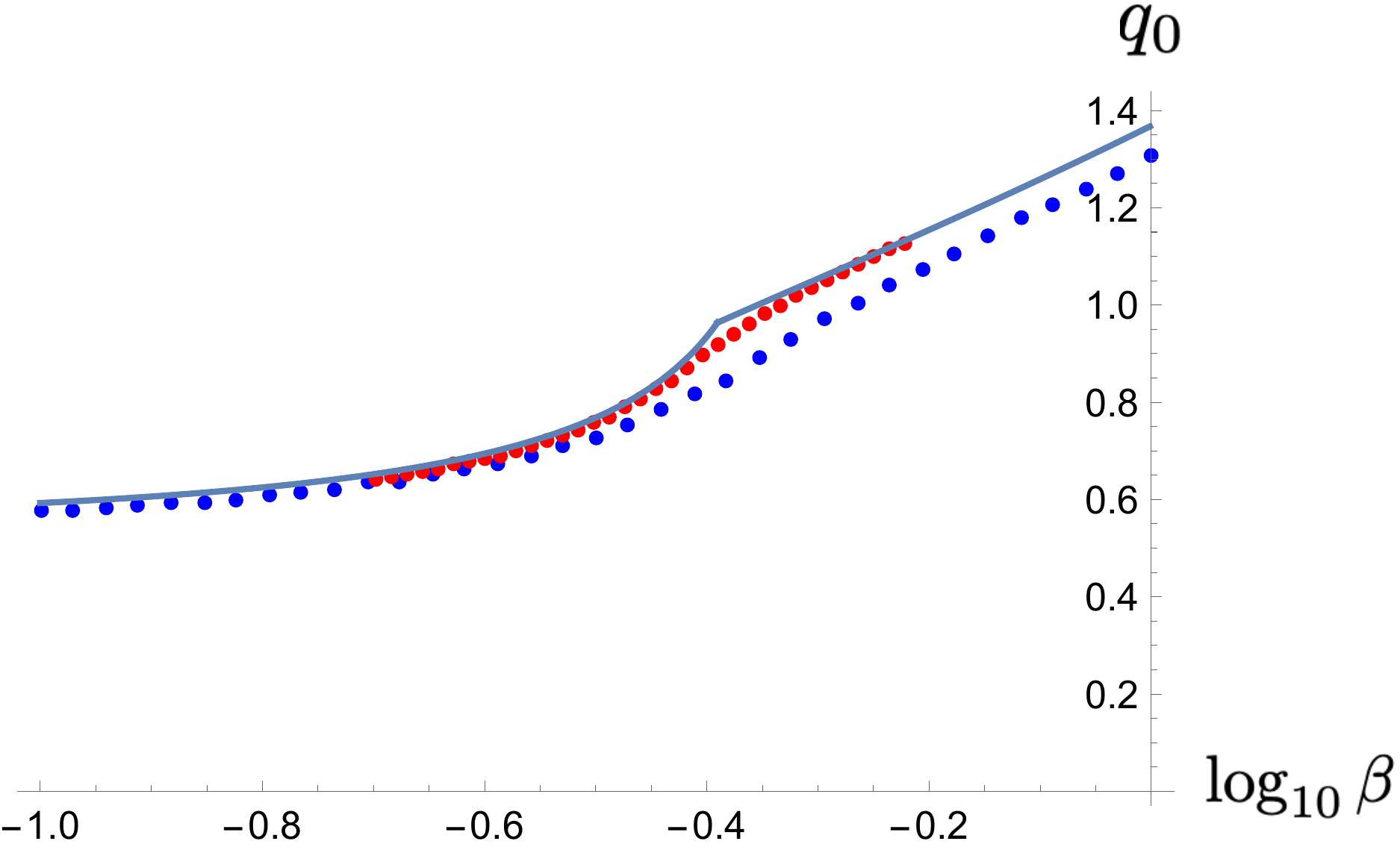}
\caption{The case with $\bar n=1,\ R=1$ with random $C$. The exact results of Section~\ref{sec:r1nbar1} are
shown by solid lines, and the numerical simulations by dots with error bars. 
$N=30$ results are shown blue, and $N=50$ shown red. 
Left two panels:  The second derivative of the free energy. 
Right panel: $q_0$ (solid line) and $(\beta/N)^{1/3} \langle \psi\cdot \psi\rangle$
(dots with error bars).}
\label{fig:nbar1}
\end{center}
\end{figure}

As for the free energy, to kill the first term in \eq{eq:freersn1} and \eq{eq:freern1}, 
which is $O(N^2)$ larger than the other terms, we compare the second
derivative of the free energy with the numerical 
results\footnote{In fact, we tried to compare the first derivative of the free energy with the simulation data
by subtracting the first term by hand. However, the data still seemed to be affected by some unknown quantities
smaller than $O(N^2)$ but significant, and we could not perform any reliable comparisons.},
which are shown in the left two panels of Figure~\ref{fig:nbar1}.
We see that the numerical results seem to converge to the exact result, though the speed of convergence is 
rather slow. Especially, the errors are very large in the right region (1RSB) of the transition point,
which is supposed to be glassy.

In the right panel of Figure~\ref{fig:nbar1}, we compare $q_0$ and 
$(\beta/N)^{1/3} \langle \psi\cdot \psi\rangle$ (Note the change of the normalization for \eq{eq:sefft}), 
and find a very good agreement between them. In this 
sense, the situation seems to be the same as in Section~\ref{sec:n2r4sim}, namely, the
numerical simulations are more reliable on configurations than on the free energy.

\subsection{$\bar n=1/2,\ R=1$, random $C$}
\begin{figure}
\begin{center}
\includegraphics[width=5cm]{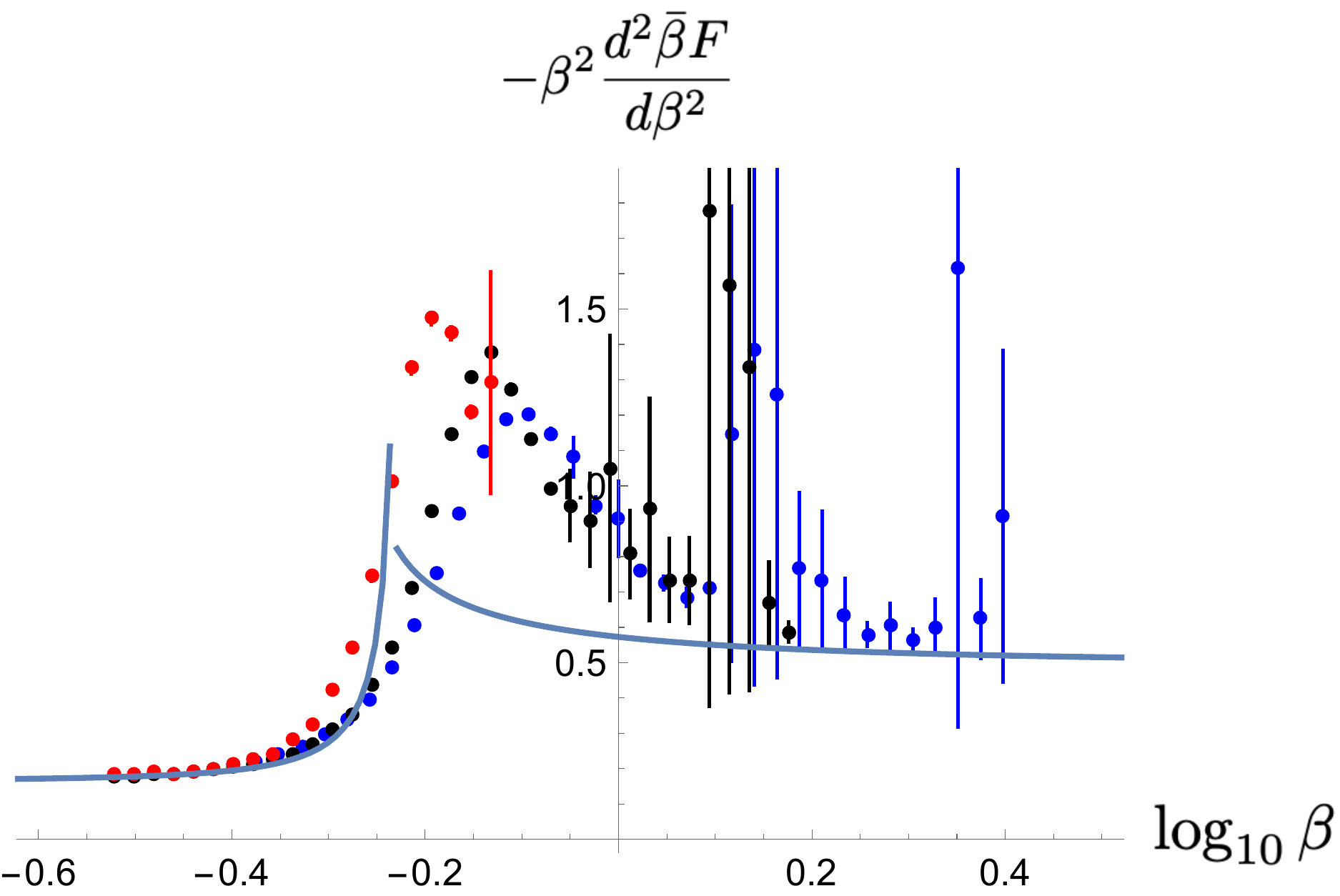}
\hfil
\includegraphics[width=5cm]{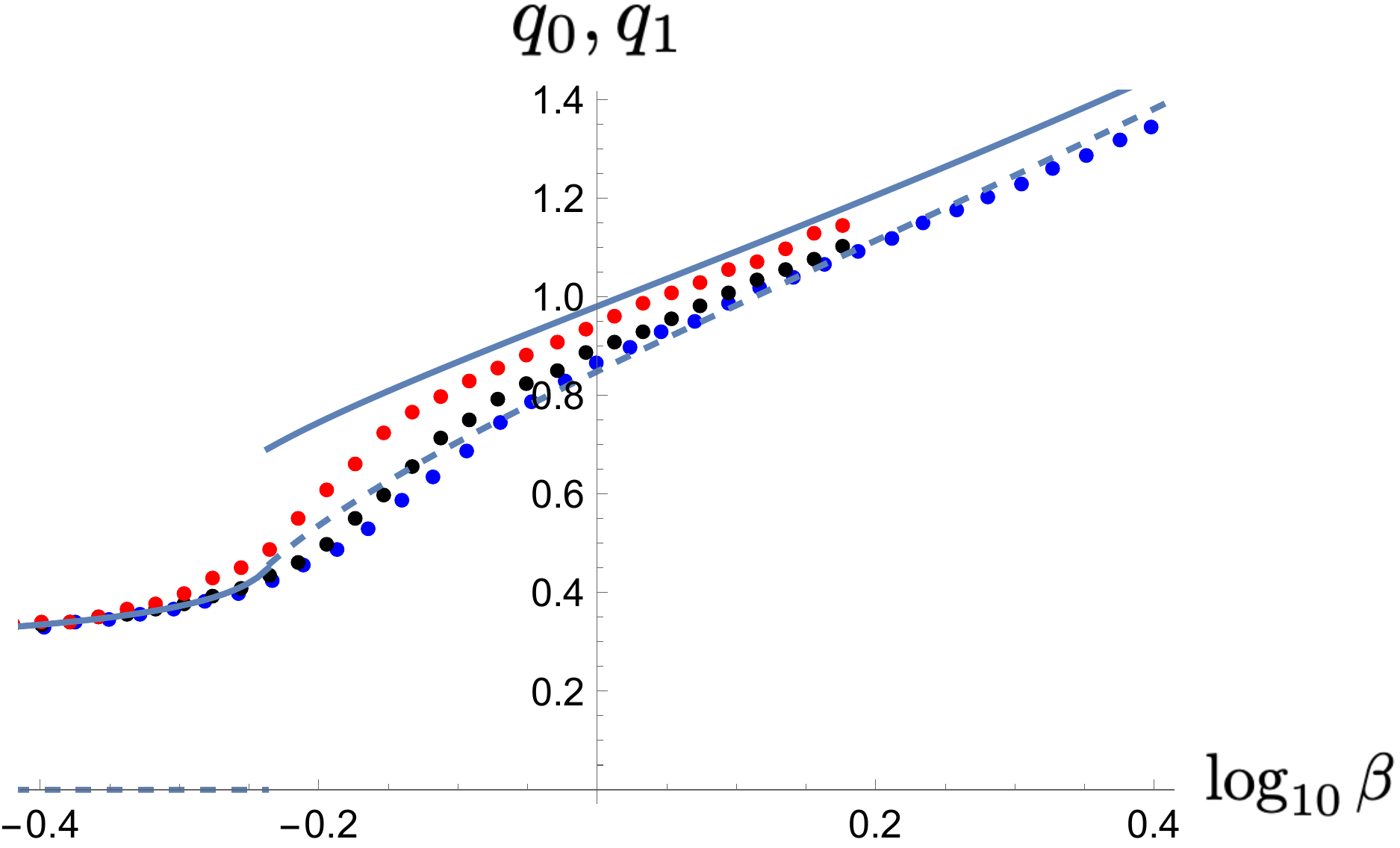}
\hfil
\includegraphics[width=5cm]{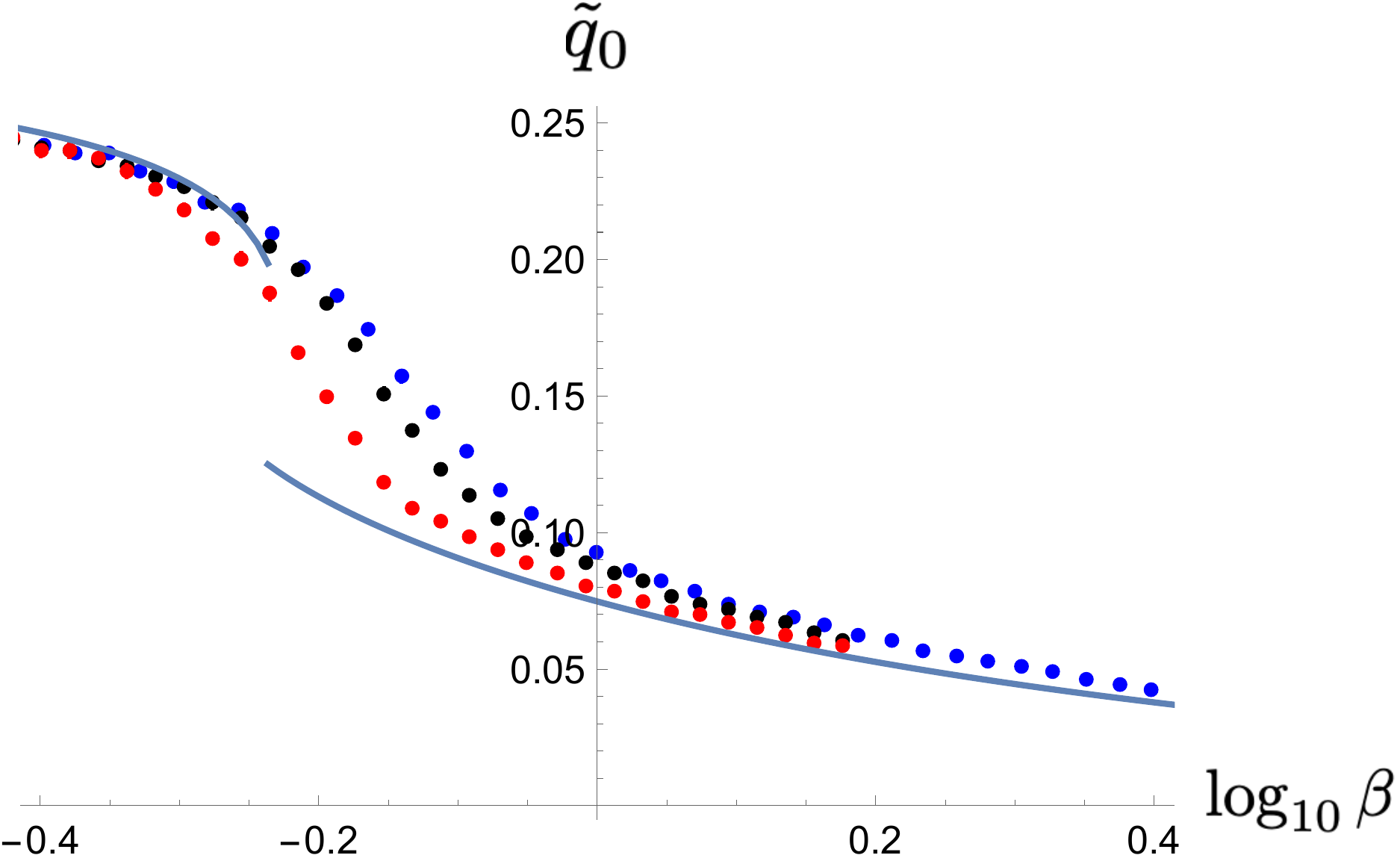}
\caption{The case with $\bar n=1/2,\ R=1$ and random $C$. The simulation data are colored: $N=30$ with blue, $N=50$
with back, and $N=70$ with red.  
Left: The second derivative of the free energy. Middle: $(\beta/N)^{1/3} \langle \psi \cdot \psi\rangle$ 
compared with $q_0$ (solid) ($q_1$ is shown dashed). 
Right: $(\beta/N)^{1/3} \langle \tilde \psi \cdot \tilde \psi\rangle$ compared with $\tilde q_0$. }
\label{fig:nhalf}
\end{center}
\end{figure}

This corresponds to the exact results in Section~\ref{sec:nhalf}.
The analysis is almost the same as in Section~\ref{sec:neq1data} but with an additional data for $\tilde q_0$.
The results are plotted in Figure~\ref{fig:nhalf}.
The convergence is much slower than the previous case.  Especially, the convergence of the second 
derivative of the free energy  is not so good, even raising some doubts about the exact results, 
but the convergence of the configurations seems to support the exact results.

\subsection{$R=2$, random $C$}
\label{sec:numnbar1rlarge}
\begin{figure}
\begin{center}
\includegraphics[width=7cm]{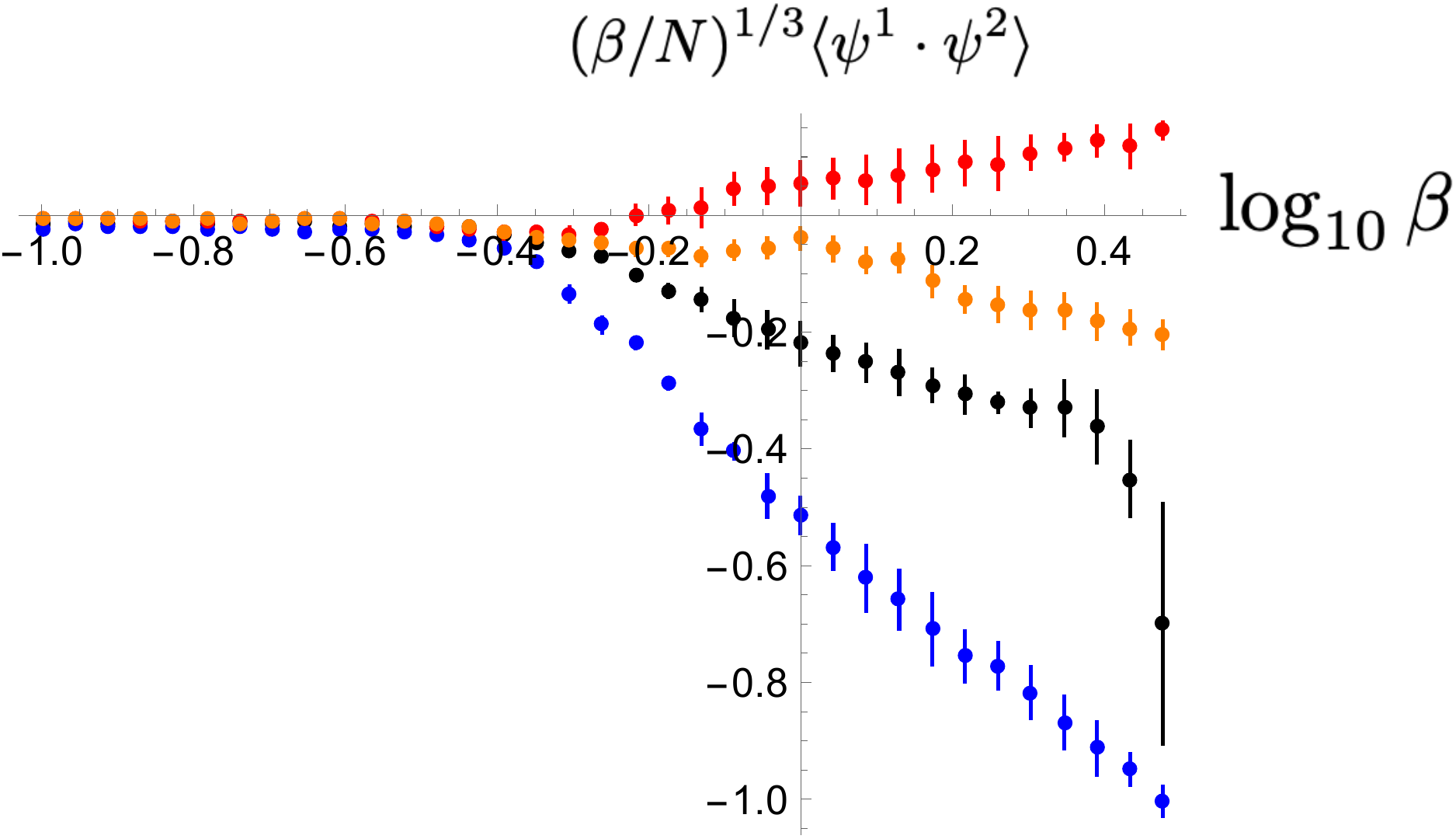}
\hfil
\includegraphics[width=7cm]{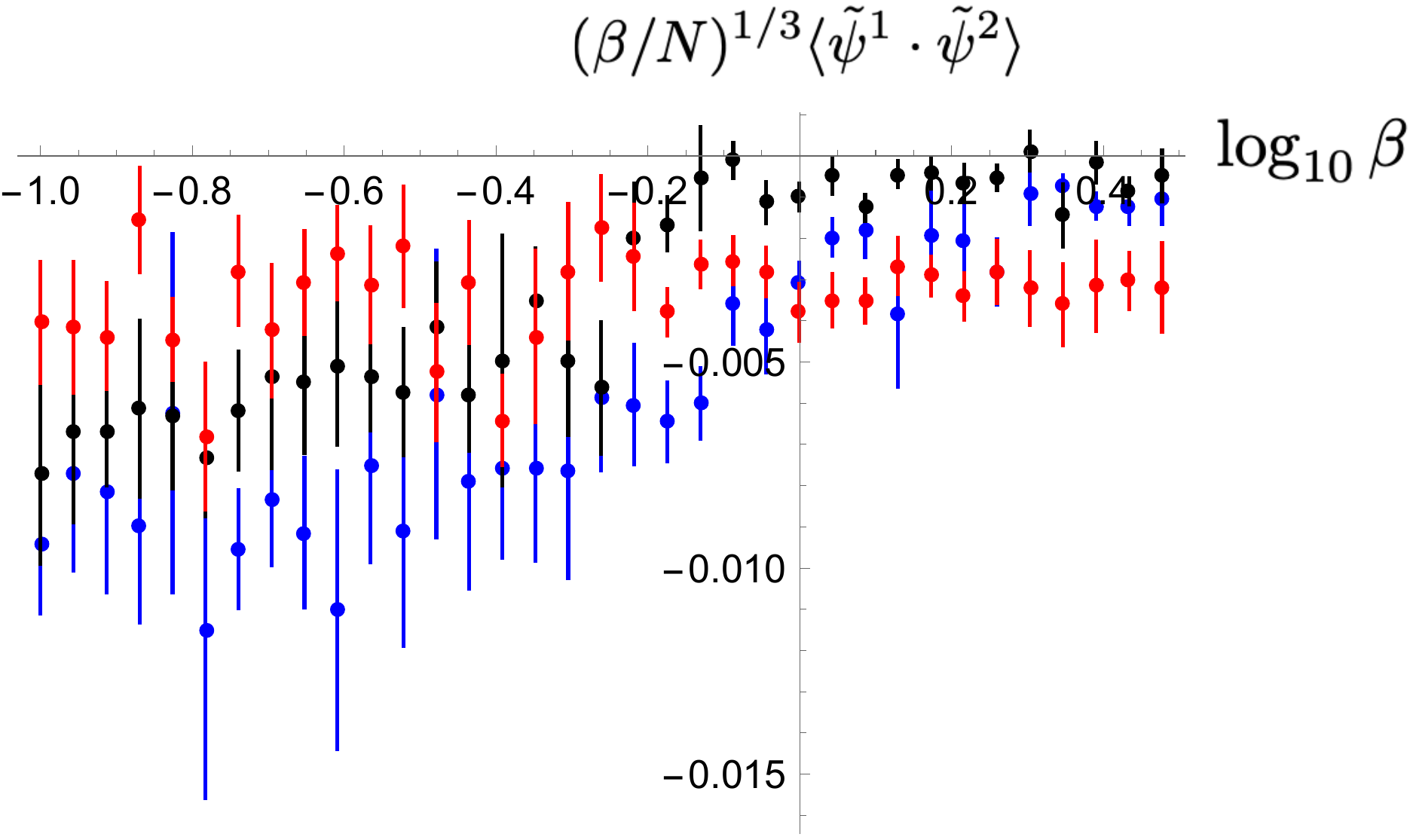}
\caption{Left panel: $\bar n=1,\ R=2$ with random $C$. 
$(\beta/N)^{1/3} \langle \psi^1\cdot \psi^2 \rangle$ is plotted for $N=30$ (blue), 
$N=50$ (black) and $N=70$ (red, orange). Two data are plotted to see the sample dependence for $N=70$.
Right panel: $\bar n=1/2,\ R=2$ with random $C$. $(\beta/N)^{1/3} \langle \tilde \psi^1\cdot \tilde \psi^2 \rangle$
is plotted for $N=30$ (blue), $N=50$ (black) and $N=70$ (red).}
\label{fig:rlarge}
\end{center}
\end{figure}

As discussed in Sections~\ref{sec:rlargenbar1} and \ref{sec:rlargensmall},
the $R>1$ cases reduce to the $R=1$ cases discussed in Sections~\ref{sec:r1nbar1} and \ref{sec:nhalf}.
We took $R=2$ and $\bar n=1,1/2$ for numerical simulations, and compared with the exact results.
As before, there are good matches for $q_0,q_1$, while the agreement on the free energies is good in the RS region, but
is difficult to compare because of large errors in the 1RSB region. 
The reduction to $R=1$ is based on that $q_0'=q_1'=\tilde q_0'=0$ obtained 
in Sections~\ref{sec:r1nbar1} and \ref{sec:nhalf}.
As shown in Figure~\ref{fig:rlarge}, $\langle \psi^1 \cdot \psi^2 \rangle$ takes non-zero
values in the 1RSB region and are dependent on samples. 
We expect that the average of these values over samples should vanish in the large $N$ limit
to match $q_0'=q_1'=0$, but the number of the data is too small to derive such a conclusion.
On the other hand, as shown in the right panel of Figure~\ref{fig:rlarge}, $\langle \tilde \psi^1 \cdot \tilde \psi^2 \rangle$
is small and would be consistent with zero, in agreement $\tilde q_0'=0$.

\section{Summary and discussions}
In matrix models, there are phase transitions in which distributions of dynamical variables change topologically,
like in Gross-Witten-Wadia transition \cite{Gross:1980he,Wadia:1980cp} 
and in the transitions among the large-$N$ limit multi-cut solutions \cite{Eynard:2016yaa}.
In a recent study \cite{Kawano:2021vqc}, 
similar splitting-merging behavior of dynamical variables
was observed in a tensor-vectors system by numerical simulations, but
the results were not convincing enough to characterize it and conclude whether this is a phase transition
or just a crossover.
In this paper, we have studied the system exactly in some large-$N$ limits,
and have found cascades of first-order phase transitions for fixed tensors, and a second- or first- order phase transition
for random tensors, applying the replica trick for the random cases.
These phases can be distinguished by breaking patterns of real replica symmetries
for fixed tensors, and those of replica symmetries for random tensors, respectively.
We have also performed some numerical simulations to compare with the exact results:
We have found consistent results, which support the assumptions made in the derivation of the exact results; 
We have also found rather slow convergence of numerical data toward sharp transitions of the exact results,
which implies the necessity of our cautious attitude toward numerical simulations of our system in future study.

As will be mentioned in the final paragraph, the large-$N$ limit of this paper is different from that in the tensor model.
However, the results of this paper suggest the following interesting possibilities.
The presence of the sharp phase transitions implies that what we called the quantum and classical phases in
the previous study \cite{Kawano:2021vqc} of the tensor model wave function could actually be different phases.  
This means that classical spacetimes could emerge through phase transitions in the tensor model. 
A new finding of this paper is that the transition rather consists of multiple first-order phase transitions. 
Therefore the classical phase, where there emerge classical spacetimes,
could not be just a single phase but rather a collection of phases.
This aspect can explicitly be seen in the $SO(2)$-invariant example given in Section~\ref{sec:lie}: 
We see that the number of discrete points forming $S^1$ increases one by one as the system undergoes 
the phase transitions, as in Figure~\ref{fig:points}. It would be straightforward to check similar matters in more 
general cases $S^n$ with $SO(n+1)$ symmetry \cite{Kawano:2021vqc}, 
and it would also be more interesting to consider general spacetimes like black holes in future study,
using the general procedure of constructing tensors 
corresponding to geometries developed in \cite{Kawano:2018pip,Kawano:2021vqc}.  

Here we would like to stress that the phase transitions in the previous paragraph are all first-order 
and should be linked solely to the emergence of classical spacetimes, which should not be mixed with 
emergence of continuum physics in them. The latter is a different thing which could be checked by 
analyzing fluctuation modes around the emergent classical spacetimes.

Another potential application of our system is to the tensor rank decomposition \cite{SAPM:SAPM192761164,Carroll1970,Landsberg2012,comon:hal-00923279}
in data analysis. 
This is an important technique to extract information from tensors of real-life data in data analysis,
but an efficient procedure is yet to be established because of its hardness \cite{nphard}. 
An interesting insight from our model is that the problems should be common across  
the tensor rank decomposition and the spin glasses, since our model can be seen bilaterally as 
a cost function of the former and a generalization of the spherical $p$-spin model for the latter.
In this sense, this paper can be seen as a single instance of using a technique developed 
in spin glass theory for a problem in data analysis (See also \cite{Ouerfelli:2022rus}). 
Though restricted to the large-$N$ limit with finite $R$ in this paper, 
the confirmation of the splitting phenomenon \eq{eq:decoupling} as a result of phase transitions seems meaningful
for the tensor rank decomposition, since the splitting gives a tensor rank decomposition with 
an automatically determined rank for an arbitrarily given tensor.
Though this phenomenon cannot be used immediately for the actual process of the tensor rank 
decomposition due to its inefficiency, 
studying our model and the phenomenon more deeply could lead to new useful procedures in the future. 

As argued in Section~\ref{sec:introduction}, our system \eq{eq:system} with random $C$ can be considered to be
a variant of the spherical $p$-spin model with a new non-trivial parameter $R$.
In this paper's limit of large-$N$ with finite $R$, however, the RS/1RSB analyses have reduced to 
the same as $R=1$, and this seemed to be consistent with the numerical results, as shown 
in Sections~\ref{sec:randomC} and \ref{sec:comparison}. While this suggests the possibility that our system is just included in the same universality class as the 
standard one, we have found large fluctuations of the overlaps between {\it real} replicas  
in Section~\ref{sec:numnbar1rlarge}. 
This fact would suggest that, for our system to be properly treated in mean field theory, we would need to 
consider many numbers of {\it real} replicas by taking large-$R$ as well. Such large-$N,R$ limits could
open new possibilities.

In fact the large-$N$ limits with finite $R$ we considered in this paper has the disadvantage that the distribution of 
dynamical variables is a finite set of configurations.
If it were continuous, the dynamics would be more interesting, like topological transitions 
as in the corresponding phenomena in matrix models.  
This disadvantage could be dealt with by simultaneously taking the large $R$ limit, which increases the number of 
overlap variables in correlation with $N$. 
In fact, there are some suggestions that urge us to look into this direction:
\begin{itemize}
\item
In \cite{Obster:2021xtb}, it was argued that the system \eq{eq:system} is convergent, only if 
$R\lesssim(N+1)(N+2)/2$. Therefore
it would be interesting to take a large-$N$ limit with $R= N^2 \, r$ with fixed $r$, and study the dependence
on $r$. Some intereresting phenomena are  expected to occur near $r=1/2$.
\item 
In the tensor model, the hermiticity of the Hamiltonian constraint requires $R=(N+2)(N+3)/4$ \cite{Sasakura:2013wza}\footnote{
As for the derivation of this value of $R$, 
it would be easier to see an appendix of \cite{Kawano:2021vqc} for a short summary with
the same normalization as is used in this paper. When the wave function is absolute-squared for probability, the value 
which matters is given by $2R=(N+2)(N+3)/2$ \cite{Sasakura:2019hql}, which curiously reproduces $r=1/2$ 
as an interesting case.}. 
This urges us to look into the same kind of limit as above.
\item
In \cite{Lionni:2019rty}, a different but closely related model was treated, and 
it was found that there is a transit parameter region, $N \lesssim R \lesssim N^2/2$,
in which dominant graphs gradually change. 
This would suggest that not only the above one but also $R =  N \, r'$ with fixed $r'$ 
could provide an interesting large-$N$ limit.
\item
From the view point of the tensor rank decomposition, 
Alexander-Hirschowitz theorem\cite{AH} tells that, with some exceptions,  
the complex symmetric general rank of order-three symmetric tensors is given by $\lceil (N+1)(N+2)/6 \rceil$, and
combining with the argument in \cite{Ballico},
the symmetric real rank of a generally given real tensor of order three 
should be smaller than or equal to $3 \lceil (N+1)(N+2)/6 \rceil$. 
This number curiously agrees well with the number given in the first item above, and gives a motivation 
to study the same large-$N$ limit.
\item 
The results in Figure~\ref{fig:rlarge} imply that $\langle \psi^1\cdot \psi^2\rangle$ 
could vanish in the 1RSB region only after the large-$N$ limit with fixed $R$ has been taken. 
Since this vanishing is the main cause for the reduction of $R>1$ to $R=1$,
we would expect more interesting dynamics to appear, if this vanishing is changed
in the simultaneous large-$N,R$ limits proposed above. 
\end{itemize}
It is stimulating that the different perspectives above actually point to the similar large-$N,R$ limits. 
We hope we can report some progress in this direction in future studies.

\vspace{.3cm}
\section*{Acknowledgements}
The work of N.S. is supported in part by JSPS KAKENHI Grant No.19K03825. 

\appendix
\def\thesection{Appendix \Alph{section}}

\section{Rank of $SO(2)$ invariant $C$}
A fact used in Section~\ref{sec:lie} is that the rank of $SO(2)$ invariant $C$ of $n=3$ is given by four. 
This is a special case of the following general formula:
\[
\hbox{Rank}(C^{SO(2)})=\frac{3n-1}{2}.
\label{eq:genrank}
\]
This statement is equivalent to that $r=(3n-1)/2$ is the minimum number 
for the following summation expression of the integral to hold:
\s[
\int_{0}^{2 \pi} d\theta f_a f_b f_c =\sum_{j=1}^r A_a^j A_b^j A_c^j, 
\label{eq:A}
\s]
where 
\[
\{f_a|a=1,2,\ldots, 2 \Lambda+1\} =\{\frac{1}{\sqrt{2}}, \cos \theta, \sin\theta, \cos 2 \theta, \cdots, \sin(\Lambda \theta)\},
\label{eq:listf}
\]
with $n=2 \Lambda +1$. 

To prove \eq{eq:A} let us recall what is called quadrature \cite{quadrature}. 
Quadrature is an approximation to an integral in terms of a summation, which however gives exact values if the integrand is 
contained in a particular set of functions.
In the present case, we want a quadrature,
\[
\int_{0}^{2 \pi} d\theta\ g(\theta) =\sum_{j=1}^r w_j\, g(\theta_j),
\label{eq:g}
\]
where $g(\theta)$ is contained in a certain set of periodic functions of period $2 \pi$, $\theta_j$ are quadrature nodes, 
and $w_j$ are quadrature weights. Let us consider the set of $g(\theta)$ to be given by
\[
\{1,\cos\theta, \sin\theta,\cos2 \theta,\ldots, \sin M \theta\}.
\] 
Then it is straightforward to show that 
the nodes can be taken as
\[
\theta_j=\frac{2 \pi j}{M+1},\  (j=0,1,\ldots,M),
\]
with weights $w_j=2 \pi/(M+1)$, because 
\[
\frac{2 \pi}{M+1} \sum_{j=0}^{M} e^{i \frac{2 \pi k j}{M+1}}=\delta_{k,0} 
=\int_0^{2 \pi} d\theta e^{i k\theta},
\]
for integer $0 \leq k \leq M$. 
Here the number of nodes, $r=M+1$, is the least, because $M+1$  independent conditions must be satisfied.

In our present case, $g$ is given by a product of three functions in \eq{eq:listf}. Therefore $M=3 \Lambda$, and 
hence $r=M+1=(3n-1)/2$, which indeed agrees with \eq{eq:genrank}. 
By applying \eq{eq:g} to \eq{eq:A}, we obtain
\[
A_a^j=\left(\frac{2 \pi}{3\Lambda+1}\right)^{\frac{1}{3}} f_a(\theta_j).
\]

\section{Details of the derivation of \eq{eq:freerlarge}}
\label{app:der}
In this appendix, we show some details of the derivation of the free energy \eq{eq:freerlarge}.

From the expression \eq{eq:qform}, we find
\s[
\sum_{ii'tt'} (\bar Q^{iti't'})^3&=R \sum_{tt'} (\bar Q^{1t1t'})^3+R(R-1) \sum_{tt'} (\bar Q^{1t2t'})^3 \\
&=RT ( q_0^3+(M-1) q_1^3) +R(R-1) T ((q_0')^3+(M-1) (q_1')^3).
\s]
We also obtain
\s[
\sum_{ii't} (\bar Q^{iti't})^3&=R \sum_{t} (\bar Q^{1t1t})^3+R(R-1) \sum_{t} (\bar Q^{1t2t})^3 \\
&=RT  q_0^3 +R(R-1) T (q_0')^3.
\s]

Computation of $\det(\bar Q)$ is also straightforward but a little involved. First we note that $\bar Q$ has the form,
\[
\bar Q^{iti't'}=A_{tt'} \delta_{ii'}+A'_{tt'} (1-\delta_{ii'}), 
\label{eq:qaa}
\] 
where 
\[
A_{tt'}=\delta_{\lfloor t/M \rfloor, \lfloor t'/M \rfloor} I_{t \bmod  M, t' \bmod  M} (q_0,q_1),
\]
and similarly for $A'$ with $q_0'$ and $q_1'$. Concerning the indices $i,i'$, 
$\bar Q$ in \eq{eq:qaa} has the ``eigenvalues", $A+(R-1)A'$
and $A-A'$ with degeneracies $1$ and $R-1$, respectively. On the other hand, $A$ has the eigenvalues, $q_0+(M-1)q_1$
and $q_0-q_1$ with degeneracies $T/M$ and $(M-1)T/M$, respectively. Combining these, we obtain the following
list of eigenvalues of $\bar Q$:
\s[
&q_0+(R-1)q_0'+(M-1)(q_1+(R-1)q_1'),\ \  \hbox{deg}=T/M, \\
&q_0+(R-1)q_0'-(q_1+(R-1)q_1'),\ \  \hbox{deg}=(M-1)T/M, \\
&q_0-q_0'+(M-1)(q_1-q_1'),\ \  \hbox{deg}=(R-1)T/M, \\
&q_0-q_0'-(q_1-q_1'),\ \  \hbox{deg}=(M-1)(R-1)T/M.
\s]
This determines the terms coming from $\det \bar Q$.

\end{document}